\documentclass[final,3p,times,twocolumn]{elsarticle}

\usepackage{amssymb}
\usepackage{siunitx}

\usepackage{lineno}

\journal{Sustainable Cities and Society}

\begin{document}
\begin{frontmatter}



\title{Invisible Walls: Exploration of Microclimate Effects on Building Energy Consumption in New York City}


\author[inst1]{Thomas R. Dougherty}
\author[inst1]{Rishee K. Jain}

\affiliation[inst1]{organization={Stanford University, Civil and Environmental Engineering},
            addressline={450 Serra Mall}, 
            city={Stanford},
            postcode={94305},
            state={CA},
            country={USA}}

\begin{abstract}
The reduction of greenhouse gases from buildings forms the cornerstone of policy to mitigate the effects of climate change. However, the automation of urban scale building energy modeling systems required to meet global urban demand has proven challenging due to the bespoke characteristics of each city. One such point of uniqueness between cities is that of urban microclimate, which may play a major role in altering the performance of energy efficiency in buildings. This research proposes a way to rapidly collect urban microclimate data through the utilization of satellite readings and climate reanalysis. We then demonstrate the potential utility of this data by composing an analysis against three years of monthly building energy consumption data from New York City. As a whole, microclimate in New York City may be responsible for large swings in urban energy consumption. We estimate that Central Park may reduce the electricity consumption of adjacent buildings by 5-10\%, while vegetation overall seems to have no appreciable impact on gas consumption. We find that favorable urban microclimates may decrease the gas consumption of some buildings in New York by 71\% while others may increase gas consumption by as much as 221\%. Additionally, microclimates may be responsible for the decrease of electricity consumption by 28.6\% in regions or increases of 77\% consumption in others. This work provides a method of curating global, high resolution microclimate data, allowing researchers to explore the invisible walls of urban microclimate which interact with the buildings around them.
\end{abstract}




\end{frontmatter}


\section{Introduction}
\label{sec:intro}
The built environment accounts for around 36\% of annual greenhouse gas emissions in Europe \cite{noauthor_focus_nodate}. In an effort to meet climate goals laid out in the Paris Agreement \cite{noauthor_paris_nodate}, urban areas around the world have adopted policies to reduce their carbon emissions. However, large scale energy modeling attempts are laden with uncertainties associated with building modeling, occupancy behavior, and socio-technical factors \cite{ali_review_2021}. The influence of urban microclimates on energy consumption has also been gaining attention in the research community, as preliminary studies show potential deviations of up to 100\% heating or 65\% due entirely to urban microclimates \cite{hong_urban_2021}.

Our study explores the potential advantages of blending high resolution climate reanalysis with remote sensing data into a statistical model of urban energy consumption for New York City. In doing so, we identify key variables and estimate their effects on regional energy consumption within the city. Finally, we use the results of the analysis to define a set of "energy microclimates" (EMC) which offer an extended definition to urban microclimates to include features most pertinent to building energy consumption and with a higher time resolution. We use these microclimate zones to then explore how buildings might shift microclimate throughout the year.

\section{Background}
Building energy modeling has been utilized as a valuable tool in the design phase of buildings for its capacity to estimate the building's energy consumption, carbon emissions, and peak electricity demand \cite{reinhart_urban_2016}. Typically, large scale energy models are partitioned by their end use case. National scale models often disregard the unique aspects of each building in favor of macroscopic grid planning, while smaller models might attempt to inject most of the pertinent information about each building through the registry of the building's archetypes in an physics based model \cite{cerezo_comparison_2017}. While both data-driven models and physics based models have shown the significance of temperature in energy consumption \cite{santamouris_impact_2015}\cite{luo_modeling_2020}, building energy models often neglect the incorporation of unique weather data. Instead, Typical Meteorological Year (TMY) data is used as a surrogate for the unique weather the building is likely to experience, which is a historical account of the regional weather data \cite{xu_better_2022}\cite{reinhart_urban_2016}. 

However, TMY data has recently been shown to poorly track localized effects like urban heat island \cite{weclawiak_viability_2022} and a lack of integration with climate models is noted as one of the major engineering obstacles to continued improvement for urban building energy modeling \cite{craig_overcoming_2022}. Recent research has motivated a new wave of interest into the topic of urban microclimates as we are beginning to understand the true implications of localized weather on our energy infratructure. In July of 2013, Toparlar et al. recorded the cooling effects from urban vegetation in Antwerp, Belgium. They then coupled their measurements with building energy simulations, discovering that buildings directly adjacent to the park had a cooling demand which was 13.9\% lower than that of the same buildings further away from the park. In their 2003 study, Kikegawa et al. estimated that the removal of waste heat from cooling systems would result in a 1${\circ}$C drop in temperature and 6\% decrease in cooling energy consumption for the region of Ootemachi, Tokyo \cite{kikegawa_development_2003}. Recent work from San Francisco used simulation to estimate the impact of microclimate on building energy consumption, finding that the failure to consider localized weather may result in up to 100\% differences in annual heating, 65\% difference in annual cooling, and up to 30\% difference in peak cooling electricity demand \cite{hong_urban_2021}. Clearly microclimate may play a significant role in driving energy consumption although it has yet to reach mainstream adoption in energy models.

The lack of microclimate integration in urban building energy modeling is likely due to limitations with modeling capacity. Modern research into microclimate weather modeling often requires the installation of sensors throughout the city with interpolation between them serving as a potential proxy for a higher resolution model of temperature \cite{hong_urban_2021}. Given that this type of urban microclimate research often involves the collection of data through the use of localized sensors, researchers often only have the capacity to isolate a single microclimate effect for analysis \cite{arens_effect_1977}\cite{chand_effect_1998}\cite{srebric_building_2015}\cite{lin_prediction_2010}\cite{li_urban_2019}\cite{yang_impact_2020}\cite{yang_contribution_2021}\cite{allegrini_influence_2015}\cite{khoshdel_nikkho_quantifying_2017}\cite{chand_effect_1998}.

Recent air quality research has demonstrated the general lack of adequate data collection of air quality in urban spaces, which may rapidly fluctuate and are potentially susceptible to localized sources of error \cite{apte_high-resolution_2017}. Given similar spatio-temporal dynamics, it is likely that urban microcliamte readings suffer from similar issues.

The capacity to study the influence of climate on urban spaces has historically been limited to the resolution of TMY data. While the past 50 years has seen a continued march of general climate models towards a higher spatial resolution \cite{sellers_global_1969}\cite{uppala_era-40_2005}\cite{hersbach_era5_2020}, the diversity of interaction effects within urban spaces mitigates the capacity to generalize the results of one study to a new region. Some preliminary studies have attempted to curate a holistic model of urban climate \cite{salvati_built_2020} or rely on a custom fluid dynamics simulation of the region to curate high resolution data \cite{toparlar_review_2017}. While custom fluid dynamics models may provide valuable insights into the heat and wind systems of a city, they are challenging to reproduce due to data curation for validation.


To address the challenges associated with high quality microclimate modeling, this work provides a potential solution for the deep integration between sources of high-resolution microclimate data and building energy modeling. Specifically, we propose the integration of high frequency satellite data with climate reanalysis. By then using this rapidly accessible microclimate data to curate an energy analysis, we aim to demonstrate the impact of microclimate on building energy usage in New York City.

\section{Data}
\begin{figure}[h!]
    \centering
    \includegraphics[width=7cm]{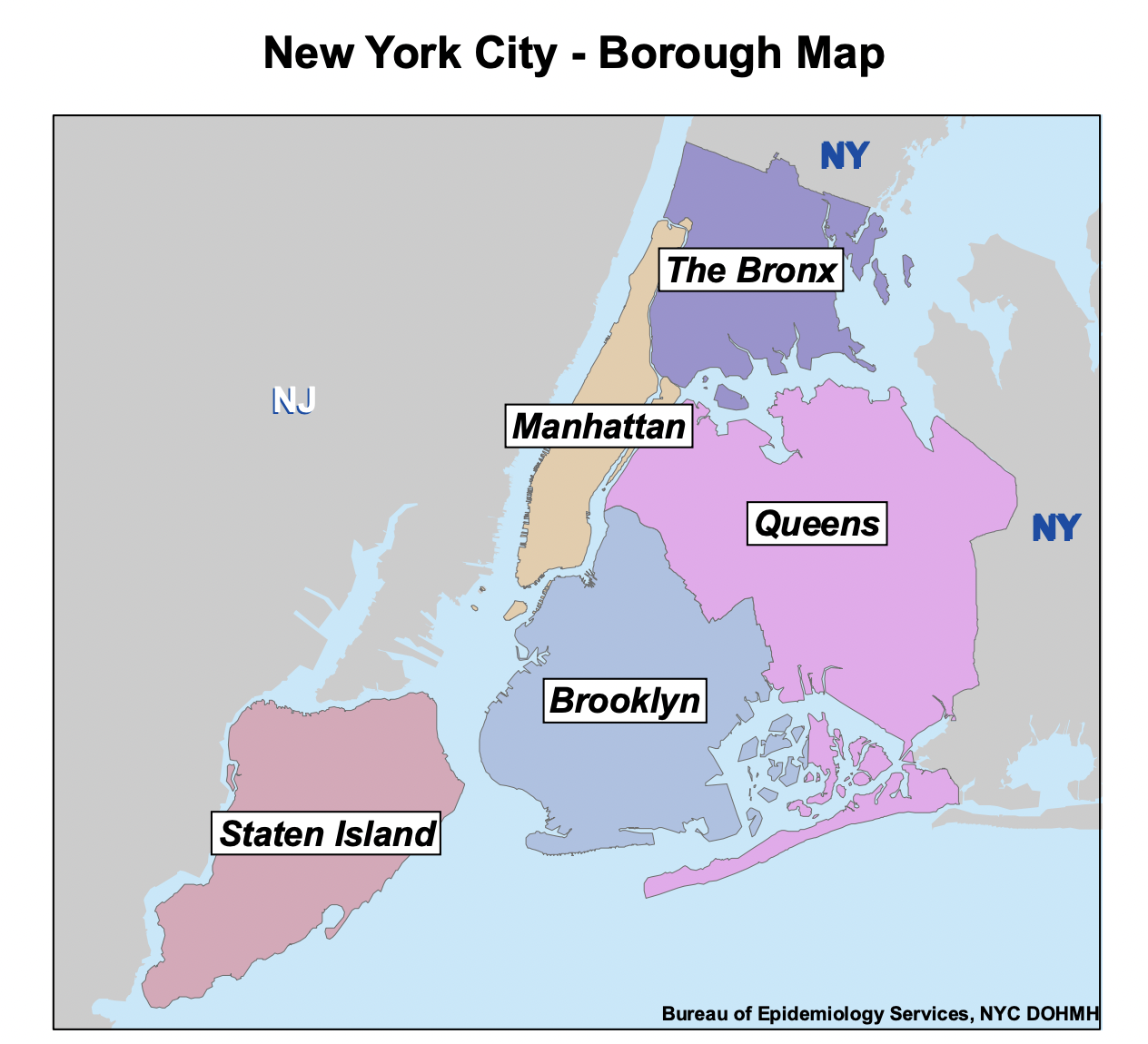}
    \caption{Overview of Boroughs in New York City}
    \label{fig:nyc_boroughs}
\end{figure}

For context, a borough map of New York City has been provided in Figure \ref{fig:nyc_boroughs} for your reference as various regions will be discussed throughout the analysis.

Our data is split into three primary categories. The first is energy data, the second is building features, and the final category is environmental data captured from satellites or reanalysis models.

All of the environmental data collected from climate reanalysis and remote sensing was captured through the utilization of Google Earth Engine \cite{gorelick_google_2017}. The process of collecting environmental parameters from Google Earth Engine is as follows. First, the convex hull of the building footprint is computed and used in place of the actual building footprint. This is due to the complexity of real building footprints, which both inflates file sizes and dramatically increases the computational complexity of subsequent analysis. Next, a buffer radius of 100 meters is applied to the building's convex hull, which will serve as the region of capture for all subsequent data. The 100 meter buffer was chosen relatively arbitrarily, with the intention to capture all potential features adjacent to the building which might have regular interactions with the structure. As per Google's documentation on the topic of image reduction \cite{noauthor_resampling_nodate}, the incoming data source will attempt to partition into the specified mesh resolution which will then be averaged over the area of the buffered region.

\begin{figure}[h!]
    \centering
    \includegraphics[width=7cm]{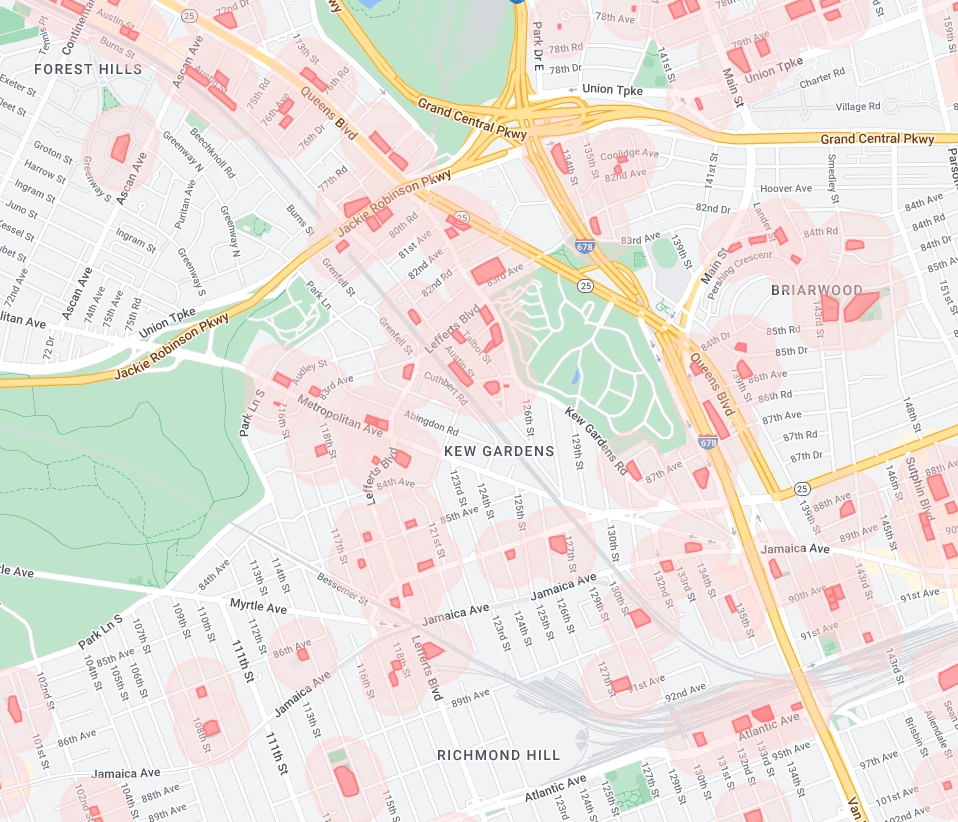}
    \caption{Subset of buffered building outlines in Brooklyn}
\end{figure}

A note on the final quality of the aggregate data set: the integration of each remote sensing instrument into the inference model brings new insights to the urban  microclimate, improving the quality of model which can be constructed. However, multiple data streams will additionally introduce new opportunities for missing data to enter the data set. In aggregate, 3.9\% of the original data is removed from the analysis due to missing data from remote sources.

\subsection{Building Data}
New York City was chosen as the test bed for this analysis due to the abundance of publicly reported data. One of the primary sources which serves as a foundation of this analysis is the monthly building energy consumption. Privately owned buildings over 25,000ft$^2$ (2,322m$^2$) and city-owned buildings over 10,000ft$^2$ (929m$^2$) are required by Local Law 84 to report their monthly electricity and gas consumption. Each building can be described with a unique ID in the form of a borough, block, and lot (BBL) number and a building identification code (BIN). This analysis links each additional source of building specific data through the building's BBL and BIN.

New York City also provides two other critical sources of information about the built environment. Building footprints for 1.08 million buildings are captured and provided by New York City's OpenData platform. The mechanism of capture for these footprints a mixture of photogrammatry and manual orthophotography, with details of edge case scenarios listed in the public disclosure of data quality provided alongside the data set.

The Primary Land Use Tax Lot Output (PLUTO) is the final New York data set used. PLUTO contains over 70 features particular to the structure, including zoning information, zip code, building use type, and wall/floor dimensions. Of note, this study intentionally leaves out a number of New York specific terms from the PLUTO data to encourage generalization to new urban regions of similar climate. The terms which are selected for the study from PLUTO are the tax value assessments (assesstot), useable building area, the year of construction (yearbuilt), and metadata used to link PLUTO data to existing data sources. The tax value metrics are captured through the Assessed Total Value and the Assessed Land Value. The usable building floor area is captured through the Building Area metric, which is procured through the collection of sources like the Property Tax System (PTS), the Computer Assisted Mass Appraisal (CAMA), or through the curation between the number of stories and building footprint \cite{noauthor_pluto_2022}.

At the time of writing in 2022, the monthly energy consumption of 9,732 buildings was reported throughout the years of 2018, 2019, and 2020. Assuming these buildings report energy data for each month, we should theoretically have 350,352 electricity data points and 350,352 gas data points. After filtering for duplicated IDs and pairing the energy consumption data of the 9,732 buildings with those of the building footprints, the total number of unique buildings falls to 9,250 with the number of valid monthly energy consumption points falling to 273,600.

\subsection{Mesoscale Reanalysis Data}
Climate reanalysis, often known through the phrase "maps without gaps", blends historical climate observations with that of modern climate models \cite{uppala_era-40_2005}. In doing so, reanalysis attempts to provide a complete picture of the earth's climate history for all locations on the planet at an hourly interval. The grid of data points provided by reanalysis is three dimensional, spanning both the surface of the planet and extending into space. As cities tend to spread more horizontally than vertically, the most pertinent measurements from climate models should be those in close proximity to the surface \cite{kondo_development_2005}.

Without much consideration to high altitude weather patterns, mesoscale models often only need to consider climate activity within the first 120 meters from the ground \cite{shi_mapping_2016}. In contrast, the mesh resolution adjacent to the surface becomes a more significant feature of interest with modern urban canopy models seeking to achieve a planar resolution of at least 1 kilometer \cite{ooka_recent_2007}. Attempts have been made to transition single layer urban canopy models to multi layer urban canopy models, but they were shown to provide no appreciable benefit for the accurate measurement of urban climate features \cite{kusaka_simple_2001}. As an illustration of scale, New York City's 99\textsuperscript{th} percentile of building heights is only 26 meters.

As such, we put more weight into selecting a climate reanalysis with high mesh resolution at the surface of the planet. This work selects NOAA's National Weather Service RTMA \cite{caldwell_real-time_nodate} for the collection of reanalysis data. The NOAA RTMA has provided historical meteorological data with a resolution of 2.5 kilometers per cell over the contiguous United States since 2011 at an hourly interval. This is the highest spatial resolution we could find among climate reanalysis models, with a data coverage which is comprehensive to our data set.

In considering alternatives, ERA5 \cite{hersbach_era5_2020} was considered for its global spatial coverage which might potentially extend this research to more regions. ERA5 was additionally found to have broad temporal reach into 1979 which would enable almost all internationally reported building energy data to be included as part of a global data set.However, ERA5's surface level resolution is roughly 30km at the equator, making it less ideal for the interpretation of microclimate causes of energy consumption. As ERA5 is often used for climate modeling research, one of the primary benefits of it is thus high resolution in the vertical direction. The lowest altitude cell is registered at 1hPa of pressure, which might corresponds roughly to 80 meters above sea level. While this is not necessarily a limiting factor in its implementation, the increased surface resolution of the NOAA reanalysis made it the more comprehensive and appropriate option.

\subsubsection{Remote Sensing}
In addition to the integration of a high resolution reanalysis model, we recognize that hyper local regional features like vegetation and impervious surfaces may have a significant influence on the energy consumption of the structure \cite{robineau_coupling_2022}. When possible, the vegetation was measured through the use of the Normalized Difference Vegetation Index (NDVI), which is a common method of locating vegetation growth based on the reflection of low frequency infrared and absorption of red light during photosynthesis \cite{noauthor_biophysical_nodate}.

Six satellite products were tested as potential candidates for this project: Landsat8, Sentinel-2 Level-2A, and Sentinel-2 Level-1C, MODIS, VIIRS, and NASA's SRTM. These satellite data sets were selected due to their prominence in scientific discourse, open access, and high resolution data. After testing, the products of Sentinel-2 Level-1C, VIIRS, and NASA's SRTM were chosen as the primary data sets for analysis.

Landsat 8 was launched on February 11, 2013 and provides global imagery at a 30 meter resolution at various wavelength intervals between 0.43 and 11.2 microns, with a revisit interval of 16 days. Landsat has been utilized for a vast set of scientific inquiries from agriculture, geology, hydrology, and the monitoring of environmental systems \cite{noauthor_applications_nodate}. Landsat provides a variety of products, with their cleanest data coming from Level 2, Collection 2, Tier 1 data. This was the data selected for analysis. The Landsat data was then masked for the presence of clouds and occlusions through the utilization of their quality bit mask, which is represented as band QA\_PIXEL. Dilated clouds, cirrus (high probability), clouds, and cloud shadows were masked. Additionally, all measurements of band saturation and terrain occlusion were masked.

Providing similar services, Sentinel-2A was launched on June 23, 2015 and Sentinel-2B was launched on March 7, 2017, the composition of which forms the Sentinel-2 constellation. The Sentinel satellites provide imagery at a much higher spatial resolution at up to 10 meters per pixel, capturing wavelengths between 0.44 and 2.2 microns with a revisit interval of 5 days. Like the Landsat satellites, Sentinel has been used in a variety of research like disaster monitoring, climate change research, and land use monitoring. \cite{phiri_sentinel-2_2020}. The raw data captured by the Sentinel satellite is processed through a series of quality control systems internal to the European Space Agency prior to releasing the data to the public. The last two phases of processing, Level-1C and Level-2A, are released for public use. They respectively provide top-of-atmosphere reflectances in cartographic geometry and bottom-of-atmosphere reflectance in cartographic geometry.

Sentinel-2 Level-2A provides a bit of additional pre-processing and added features compared to Sentinel-2 Level-1C, initially making it a more attractive choice. However, we found that the collection of imagery through Google Earth Engine for the year 2018 was unavailable over New York City, rendering a large fraction of our data set useless. Initial explorations found that Sentinel-2 Level-1C seemed to still meet our needs well and the cloud masking provided with the Level-1C product was adequate in performance without distorting the quality of the data collected.

The cloud masking for both sentinel products was procured through the provided bitwise mask band. For Sentinel-2 Level-1C this was the QA60 band, while for Sentinel-2 Level-2A this was the scene classification level (SCL) band. The Level-1C product was masked for clouds and cirrus, while the Level-2A product was masked for medium probability clouds, cloud shadows, and dark areas.

\begin{figure}[h!]
    \centering
    \includegraphics[width=7.5cm]{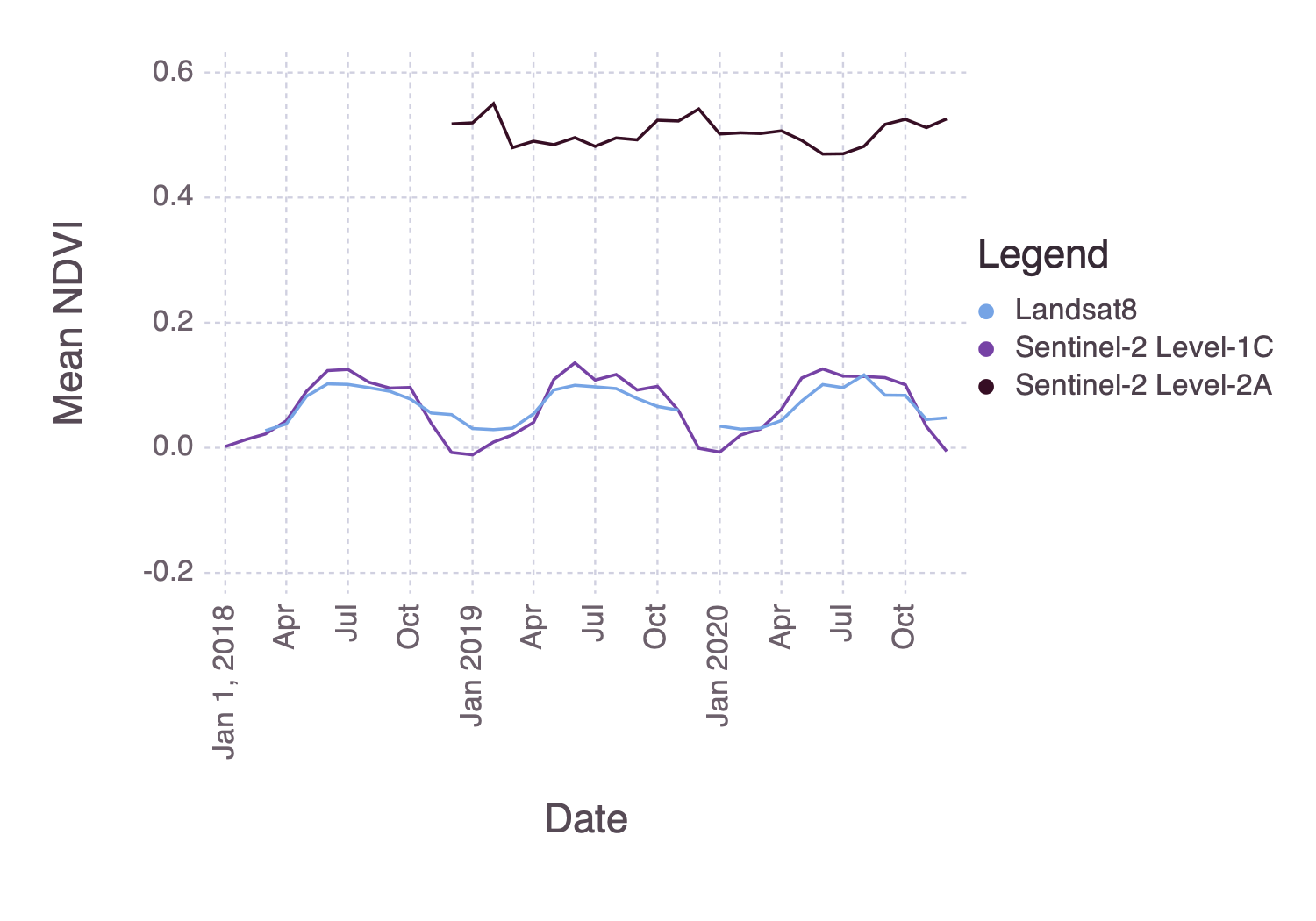}
    \caption{Satellite Comparison - Mean NDVI}
\end{figure}


While the likelihood of microclimate measurement adjacent to buildings was uniformly increased through the use of the Sentinel2-Level 1C data, the most significant improvements came in high density urban regions in Brooklyn, Queens, upper Manhattan, and Staten Island.

\begin{figure}[h!]
  \centering
  \includegraphics[width=7cm]{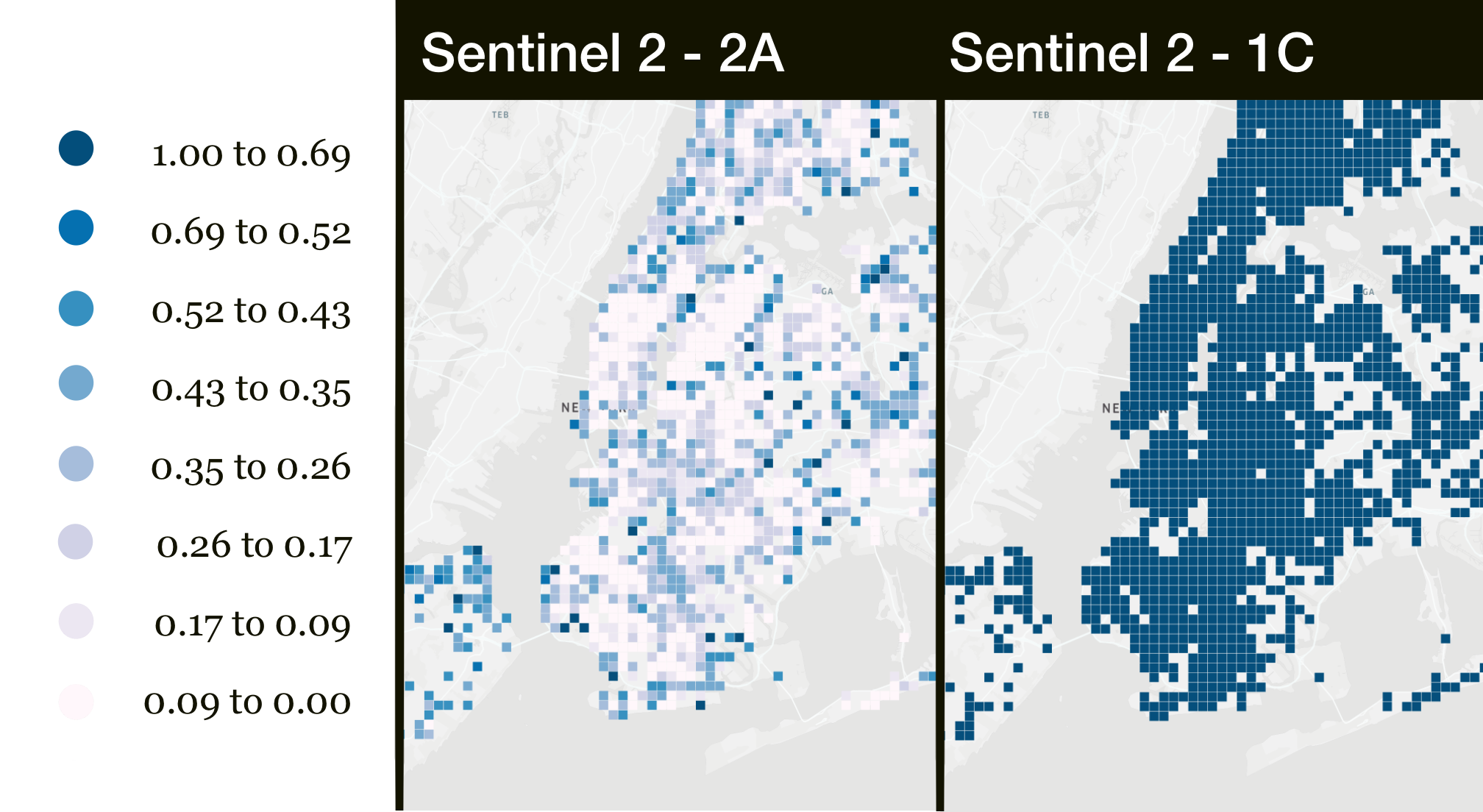}
  \caption{Percent of Months Recorded}
  \label{fig:sentinel_accuracy}
\end{figure}

While similar patterns were recorded, Landsat8 was found to consistently under report NDVI compared to Sentinel2-Level 1C. Likewise, a handful of months with particularly large discrepancies was the cause of significance difference between the measurements. Landsat8 differed from Sentinel2-Level 1C NDVI measurements with an average percent difference of 183\%. Nonetheless, the shape of NDVI readings from Sentinel2 Level-1C seemed to reasonably match that of Landsat 8. Because of its similar behavior, superior spatial resolution, and increased revisit frequency, the Sentinel2-Level 1C product was selected as the primary source of remote sensing data for the remainder of the study.

\begin{figure}[h!]
  \centering
  \includegraphics[width=5.5cm]{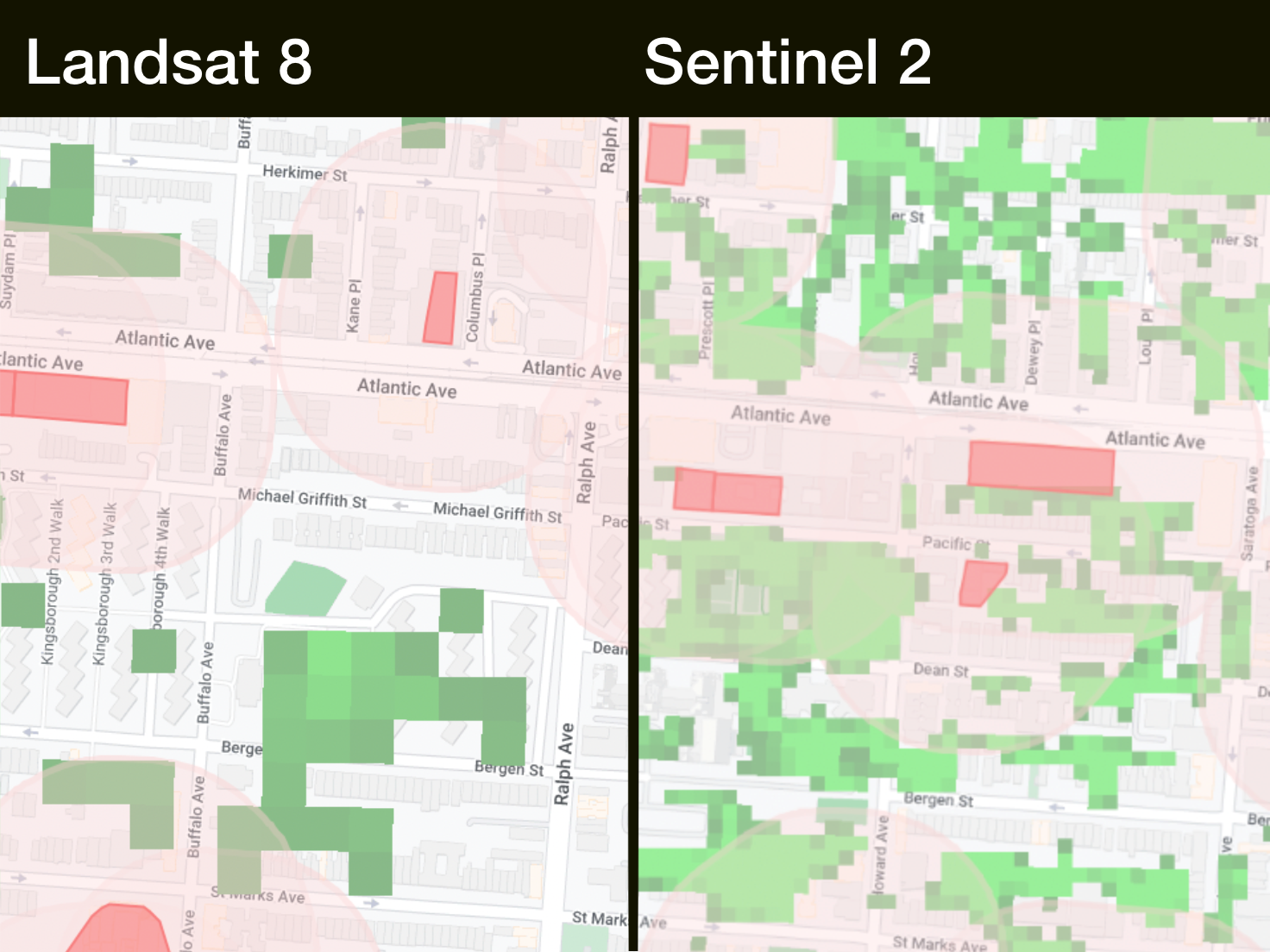}
  \caption{NDVI Capture Resolution}
  \label{fig:ndvi_difference}
\end{figure}




Land surface temperature (LST) measurements were also collected from the MODIS constellations, which were launched in 1999 and 2002. The two MODIS satellites have a relatively poor spatial resolution of 1.2km but benefit from high temporal resolution with revisit times of 2 days. When queried from Google Earth Engine, downtown Manhattan was noticeably absent for all time periods of collection. This hole in the data excludes a large fraction of our energy data from the analysis. In addition, coastal regions seem to have large gaps in collection, potentially making future analysis for other coastal cities less practical. For these reasons, MODIS data was not selected for further analysis.

Urban nightlights were the next source of exploration in remote sensing data as they are shown to provide insights into the spatial connectivity of urban spaces \cite{small_mapping_2013} and hints of economic activity \cite{dasgupta_using_2022}\cite{maatta_nighttime_2021}. Our hypothesis is that urban areas saturated with lighting might be indicated as commercial zones, which may serve as a valuable feature for prediction. The night lights data was provided by the Earth Observation Group at the Payne Institute for Public Policy, Colorado School of Mines \cite{elvidge_viirs_2017}, capturing radiance using the VIIRS instrument on the Joint Polar Satellite System (JPSS) between the wavelengths of \SI{412}{\nano\metre} and \SI{12}{\micro\metre}.

Finally, ground elevation data was folded into the analysis through the incorporation of NASA's Shuttle Radar Topography Mission (SRMT) in 2000. The static map created by this endeavor provides a 30 meter resolution map of global elevation.

\begin{figure}[h!]
    \centering
    \includegraphics[width=5cm]{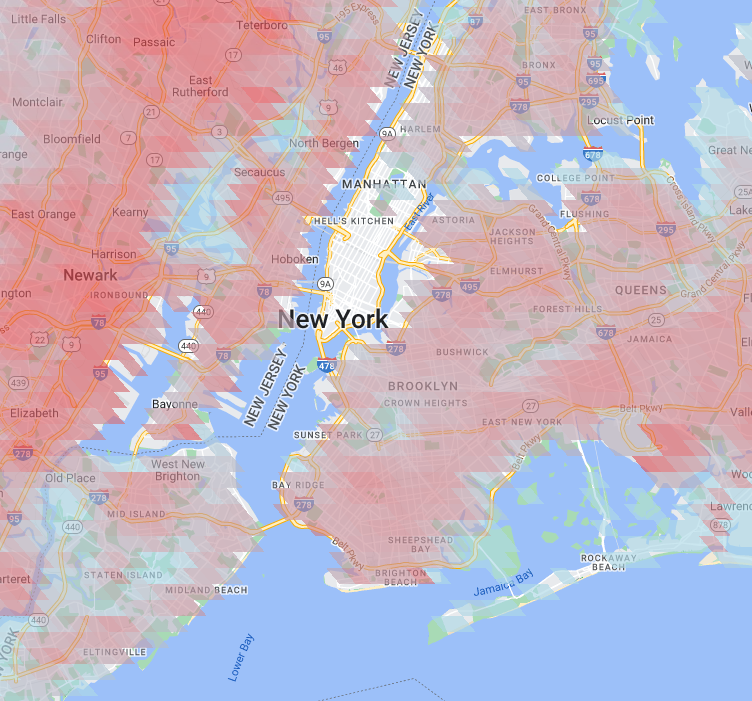}
    \caption{MODIS median daytime LST: July 26 - September 5, 2019}
\end{figure}

\subsubsection{Cloud Coverage and Radiance Quality}
As with global climate models, clouds provide a significant source of uncertainty and bias in the data we collect. Prior work has shown that the mean annual cloud coverage over land in the mid-latitudes is approximately 35\% \cite{ju_availability_2008}. When thinking instead about global land cloud coverage, this annual statistic jumps to between 58\%-66\% \cite{rossow_advances_1999}\cite{zhang_calculation_2004}. To mitigate the influence of clouds and maximize usable data in this work, we apply cloud masks to the images prior to their incorporation in the data pipeline. This enables our work to control for uncertainties associated with cloud coverage and radiometric saturation.

The cloud mask is composed by identifying high B2 (blue light) reflectance (490nm). To avoid the false detection of snow,  short-wave infrared reflectance (1610nm - 2200nm) is incorporated, which has high values for clouds and low values for snow. Cirrus clouds are semi-transparent, so their reflectance of the B2 band is much lower. For cirrus cloud masking, high reflectance along the B10 (1375nm) band is matched to low reflectance in B2 \cite{noauthor_level-1c_nodate}. These masks are provided as part of the Sentinel-2 Level-1C product at a resolution of 60 meters per pixel.

\begin{figure}[h!]
    \centering
    \includegraphics[width=6.5cm]{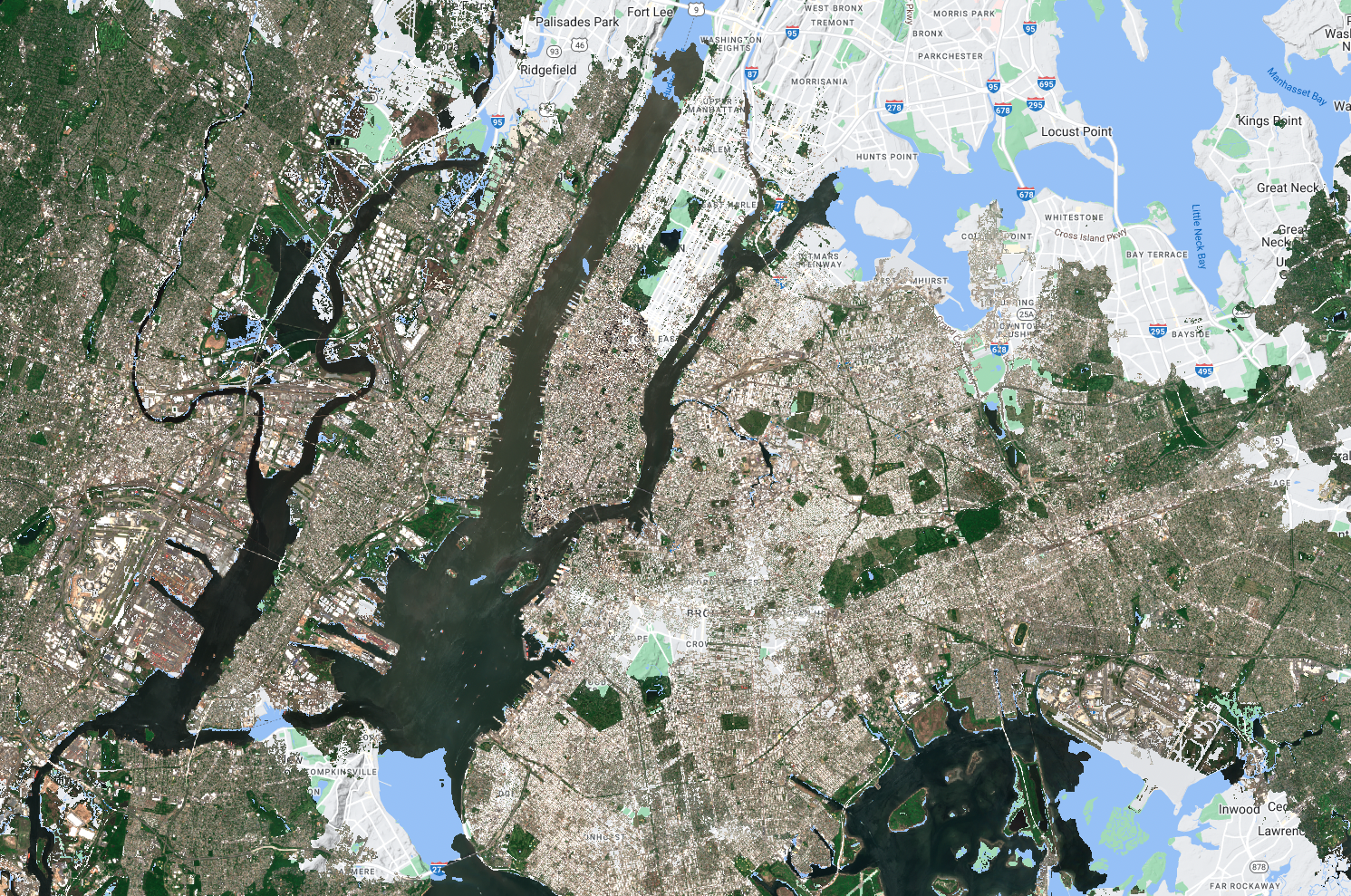}
    \caption{Sample Sentinel-2 Level-2A Image with Quality Mask: May 19, 2019, 15:51:59}
    \label{fig:cloud_mask_quality}
\end{figure}

\begin{figure}[h!]
  \centering
  \includegraphics[width=6.5cm]{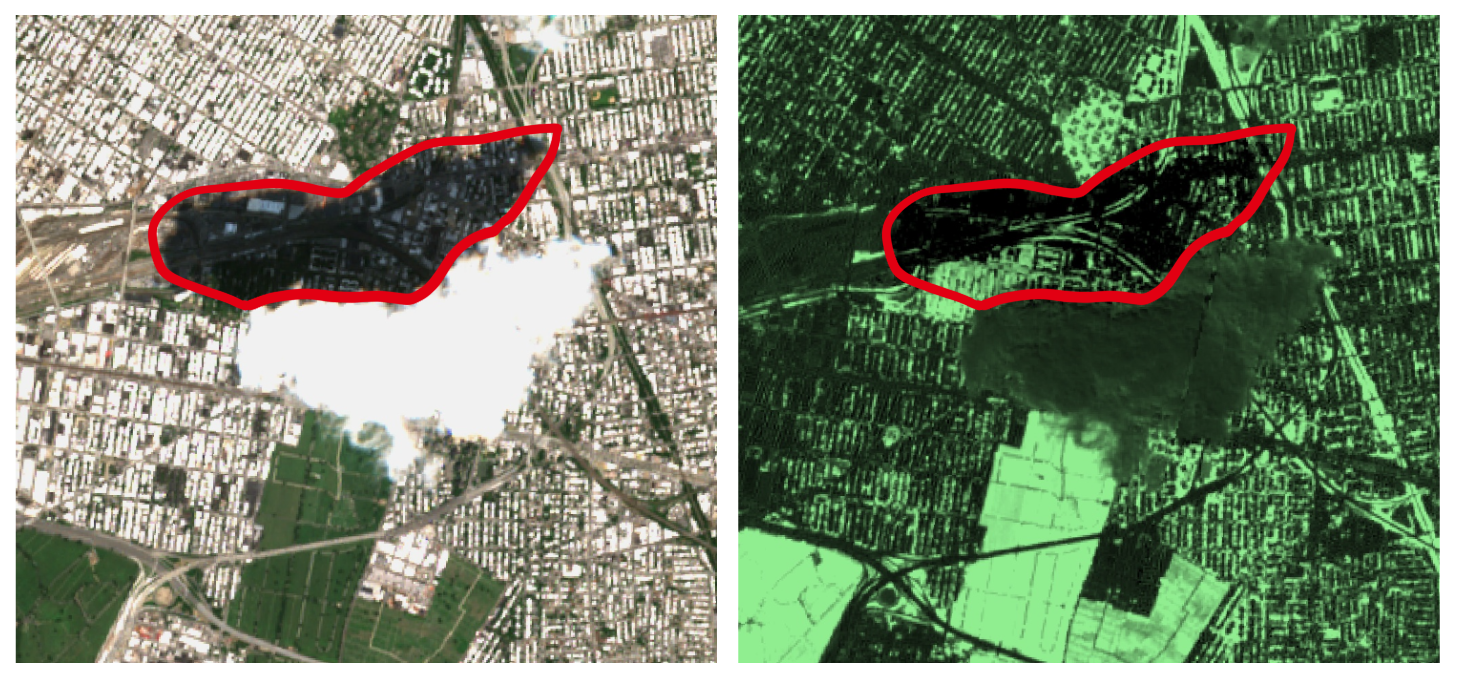}
  \caption{Cloud Shadow Impact on NDVI readings - Queens}
  \label{fig:cloud_ndvi}
\end{figure}

The cloud masking system of Landsat and Sentinel2-Level 1C seem to yield similar results while Sentinel2-Level 2A outputs a trend which is both inverted and enlarged. Conventional thinking would imply that the NDVI should increase during the summer months and decrease in the winter. Given that the algorithm used during the study of Sentinel-2A was more aggressive at masking cloud shadows, it might be possible that the shadows from dense urban areas might be mistaken for cloud shadows and filtered from the dataset. The aggressive cloud masking additionally poses a potential issue for data quality, as seen in figure \ref{fig:cloud_mask_quality}. Within the Sentinel2-Level 2A data set, a typical building has a 2.6\% chance of being located in any given photo. This probability increases to 14.0\% for the Sentinel2-Level 1A data. Another point of note is that it doesn't seem the cloud shadows have a significant impact on the capacity of the satellite to detect NDVI, as evidenced by a sample image and subsequent NDVI mask in Figure \ref{fig:cloud_ndvi}.

\section{Methods}
Having conducted a general survey of potential climate models and satellite data, we proceed with analysis through the use of linear regression against the final variables found in table \ref{tab:data_summary}.

\subsection{Endogenous Terms}
The endogenous terms ($Y$) are defined by the equation $Y_i = ln(E_i/A_i)$, with $i \in B$ and $B$ defined by the set of all buildings in the data set. Two variables are selected as the endogenous terms ($E$): the monthly electricity consumption and the  monthly gas consumption, both in units of MWh. These terms are divided by the useable floor area ($A$) of the buildings ($B$), which is provided by the PLUTO dataset as sq.ft and translated to square meters. A logarithm is then applied to this area normalized energy consumption, which both reduces the influence of right skew and permits analysis of each variable's effect size per unit increase in the term.

\subsection{Exogenous Terms}
\begin{figure}[h!]
    \centering
    \includegraphics[width=7cm]{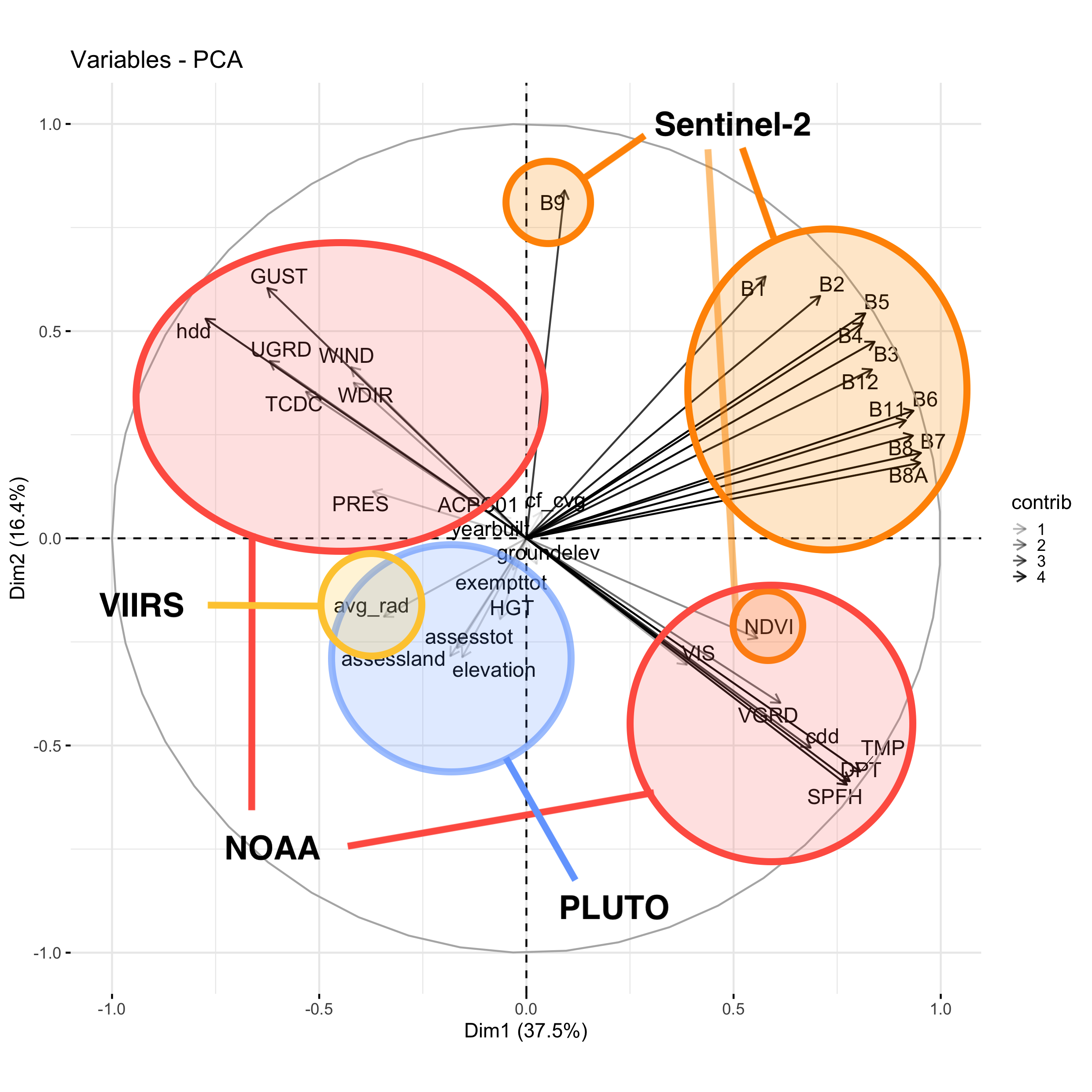}
    \caption{PCA - Potential Features}
    \label{fig:correlation_matrix}
\end{figure}

Of particular interest for this inference-based study is the issue of multicollinearity, which may cause the coefficients of our regression parameters to wildly fluctuate if the exogenous terms are too highly correlated. The terms are evaluated by computing their variance inflation factors (VIF), which provide an indication for the degree of multicollinearity by quantifying the capacity of the exogenous terms to act as predictors for a single other term \cite{noauthor_107_nodate}. The plot found in Figure \ref{fig:correlation_matrix} is a correlation matrix of the exogenous variables, which are sequentially removed until all VIF terms are below 5.


A single feature from each of these highly related groups was used to represent the entirety of the correlated terms. B1 - B12, with the exception of B11, were represented by the single band of B1. The specific humidity and dewpoint temperature were represented using heating degree days and cooling degree days \cite{damico_building_2019}, which was computed using the temperature measurements from the NOAA reanalysis. Gust, directional wind speeds, pressure, and visibility were all highly correlated with wind speed and thus removed. Finally, land value was removed for its high correlation with total property value in addition to redundant elevation terms.

The final variables selected, the data set of origin, and its computed VIF value can be found in Table \ref{tab:data_summary}.

\begin{table*}[ht]
\caption{Final Variables Description}
\centering
\begin{tabular}{rllrrlr}
  \hline
 & Variable & Description & Unit & Resolution & Data Source & VIF \\ 
  \hline
  1 & avg\_rad & Nighttime Light Radiance & \SI{}{\nano\watt \cdot \steradian^{-1} \cdot \centi\metre^{-2}} & \SI{500}{\metre} & JPSS VIIRS/DNB & 1.45 \\ 
  2 & B1 & Aerosols (\SI{443}{\nano\metre}) & \SI{}{\watt \cdot \steradian^{-1} \cdot \metre^{-2}} & \SI{60}{\metre} & Sentinel-2 Level-1C & 2.61 \\ 
  3 & B11 & Shortwave Infrared (\SI{1612}{\nano\metre}) & \SI{}{\watt \cdot \steradian^{-1} \cdot \metre^{-2}} & \SI{20}{\metre} & Sentinel-2 Level-1C & 4.00 \\
  4 & NDVI & Vegetation Index & - & \SI{10}{\metre} & Sentinel-2 Level-1C  & 2.09 \\ 
  5 & WIND & Average Wind Speed & \SI{}{\metre \cdot \second^{-1}} & \SI{2.5}{\kilo\metre} & NOAA RTMA & 1.60 \\ 
  6 & TCDC & Total Cloud Cover & \% & - & NOAA RTMA & 2.81 \\ 
  7 & ACPC01 & Total Precipitation & \SI{}{\kilo\gram \cdot \metre^{-2}} & \SI{2.5}{\kilo\metre} & NOAA RTMA & 1.60 \\ 
  8 & hdd & Heating Degree Days & - & \SI{2.5}{\kilo\metre} & NOAA RTMA & 3.37 \\ 
  9 & cdd & Cooling Degree Days & - & \SI{2.5}{\kilo\metre} & NOAA RTMA & 3.23 \\ 
  10 & elevation & Elevation & \SI{}{\metre} & \SI{30}{\metre} & NASA SRTM & 1.16 \\ 
  11 & assesstot & Total Property Value & \$ & - & PLUTO & 1.15 \\ 
  12 & yearbuilt & Year of Construction & - & - & PLUTO & 1.03 \\ 
  \hline
\end{tabular}
\label{tab:data_summary}
\end{table*}

\subsection{Regression Construction}
For each regression constructed, two variations are proposed. The first analysis is composed by regressing the endogenous energy consumption against mean centered exogenous terms. By choosing not to normalize unit variance prior to regression, we can estimate the percentage change on energy per unit area given a unit increase in the exogenous terms. The second regression is instead constructed by both mean centering and normalizing the exogenous terms, with the intention of interpreting the magnitude of significance for the term, given that some of the features might have greater variance throughout the city.

\subsection{Regression and Microclimate Identification}
\label{regression_description}
After the removal of all zero terms from the endogenous variable and corresponding exogenous features, relationships are discovered through the use of linear regression. The results of the mean centered regression can be found in Table \ref{tab:regression_mean_centered}, while the results of the normalized regression can be found in Table \ref{tab:regression_normalized}. To help identify microclimate regions within the city, the regression is utilized as a point of information for subsequent clustering.

Typical linear regression looks to minimize the sum of mean squared loss $L(\beta) = || \textbf{X}\beta - \textbf{Y} ||^2$. For an array of samples, $\textbf{X}\beta$ thus becomes a matrix-vector product. The dimensionality of $\textbf{X}$ is $NxM$, where $N$ is the number of samples while $M$ is the number of features used to describe each building. As $\beta$ represents the coefficients from the regression its dimensionality is $Mx1$, and thus the matrix vector product results in the prediction $p_i = \sum_{j=1}^{M} \textbf{X}_{ij} \beta_j$. If instead the summation is removed and an augmented matrix $\textbf{A}$ of dimensionality $NxM$ is created such that $\textbf{A}_{ij} = \textbf{X}_{ij}\beta_j$, then clustering against $\textbf{A}$ will uncover microclimate communities which are particularly significant for energy consumption. Two $\textbf{A}$ matrices will be constructed, one for gas - $\textbf{A}_g$ and one for electricity $\textbf{A}_e$.

As environmental parameters are often linked by underlying land use changes, it may be unlikely that a single environmental variable will ever be shifted without modification to others. Thus for a more accurate simulation of potential modifications to urban microclimate, we propose that shifts between clusters may instead be simulated. For this analysis, gaussian mixture models were explored as potential clustering mechanics for $\textbf{A}$.




\section{Results and Discussion}
The results of the regression may be found in Tables \ref{tab:regression_mean_centered} and \ref{tab:regression_normalized}. Our results indicate that localized microclimate indeed seems to play a significant role in urban energy consumption, and many of the pertinent microclimate features may be captured using the resolution of data provided by modern climate models and satellite imagery.

\subsection{Electricity}
The microclimate causes of electricity consumption in New York seem to be relatively heterogeneous, as seen in the regression results from Table \ref{tab:regression_normalized}. The most significant microclimate features to drive consumption habits seem to be the cooling degree days, heating degree days, and B11 readings. B11 captures light at a wavelength of 1,610 nanometers, which is classified as shortwave infrared radiation. An increase in measured values of shortwave radiation might indicate that the material is able to dissipate thermal energy, whereas lower readings might indicate a level of thermal trapping which may be caused by the urban canopy.

Night light emissions (avg\_rad) are also identified as a significant term in the regression model. While this could be due to the raw energy consumption of lighting or perhaps more likely, higher levels of night time emissions may be a statistical artifact of commercial regions in the city.

We may also use these results to explore localized microclimate effects and estimate the influence of various urban features. For example, with a coefficient of -0.754, the mean centered regression indicates that an increase of one unit in NDVI would correspond to a decrease of 75\% in electricity consumption. In practice, the typical  NDVI in New York City might swing between -0.05 and 0.2 depending on the season and density of vegetation in the region. As an example, the average NDVI values in Midtown (middle of Manhattan) are -0.04. With a regression coefficient of -0.754, NDVI is expected to modify the electricity consumption in Midtown by $100 \cdot (exp(-0.754 \cdot -0.04) - 1) = 3.06\%$.

Buildings adjacent to Central Park on the Upper West Side (northern Manhattan) have average readings closer to 0.11. The estimated impact from vegetation on electricity is thus $100 \cdot (exp(-0.754 \cdot 0.11) - 1) = -7.96\%$. This indicates that an identical building next to Central Park on the Upper West Side is likely to have 11\% lower electricity consumption than in Midtown due to the difference in vegetation, which is similar to the results found by Toparlar et al. in their analysis of cooling demand in Antwerp, Belgium \cite{toparlar_impact_2018}. The same method of computation may be applied to any of the regression coefficients in Table \ref{tab:regression_mean_centered} to estimate the feature's impact on the endogenous term.

\subsection{Spatial Analysis - Electricity}
\begin{figure}[h!]
    \centering
    \includegraphics[width=6.5cm]{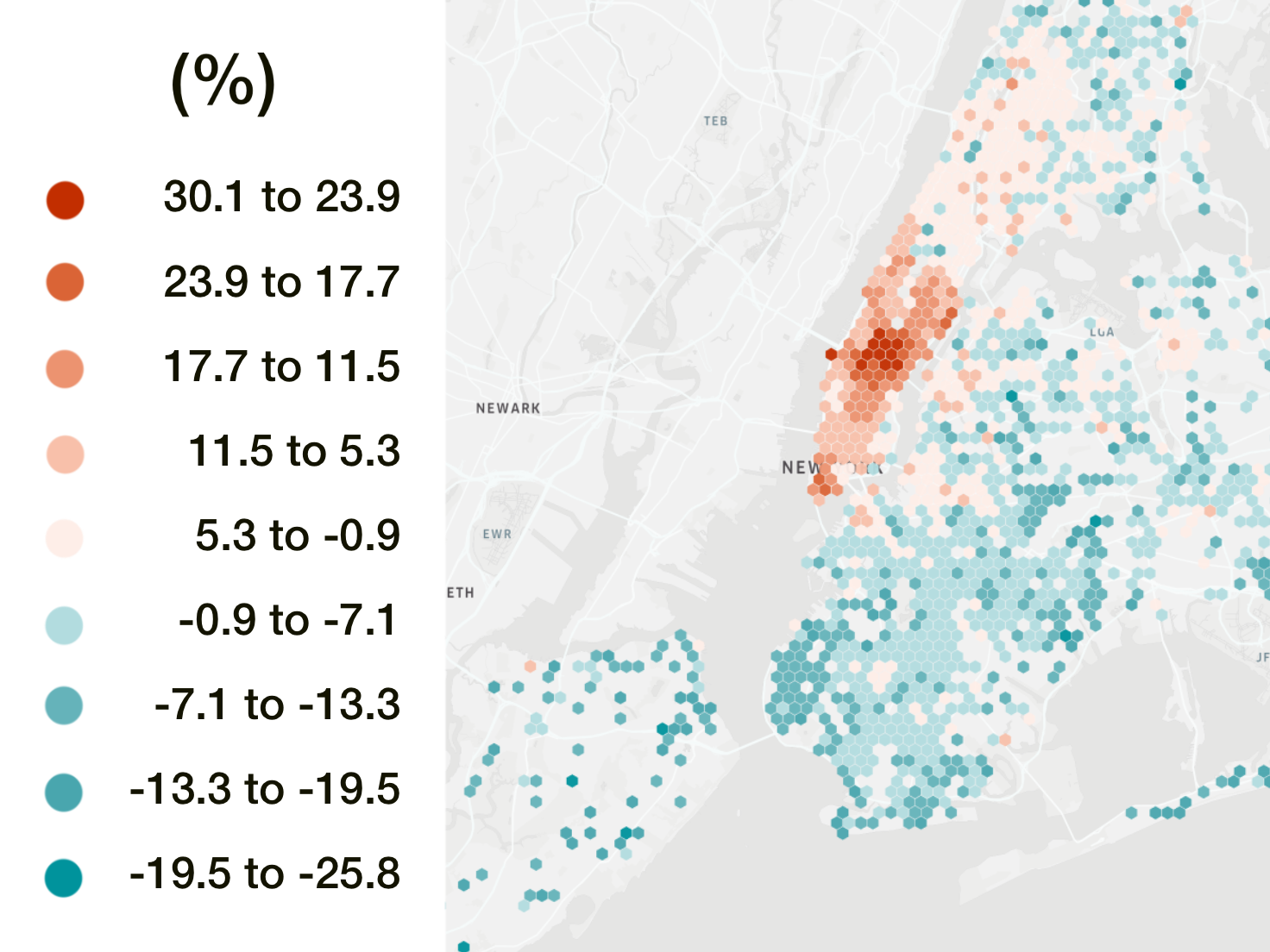}
    \caption{Environmental Percent Difference - Electricity}
    \label{fig:environmental_electricity}
\end{figure}

In general, the more densely packed urban areas in New York seem to suffer from negative effects of urban microclimate with regards to electricity consumption. The lower B11 readings in Manhattan indicate that trapping of low-frequency light might be disproportionately impacting the efficiency of the structures. The overall estimated impact of environmental features on electricity consumption is shown in Figure \ref{fig:environmental_electricity}.

\begin{figure}[h!]
    \centering
    \includegraphics[width=6.5cm]{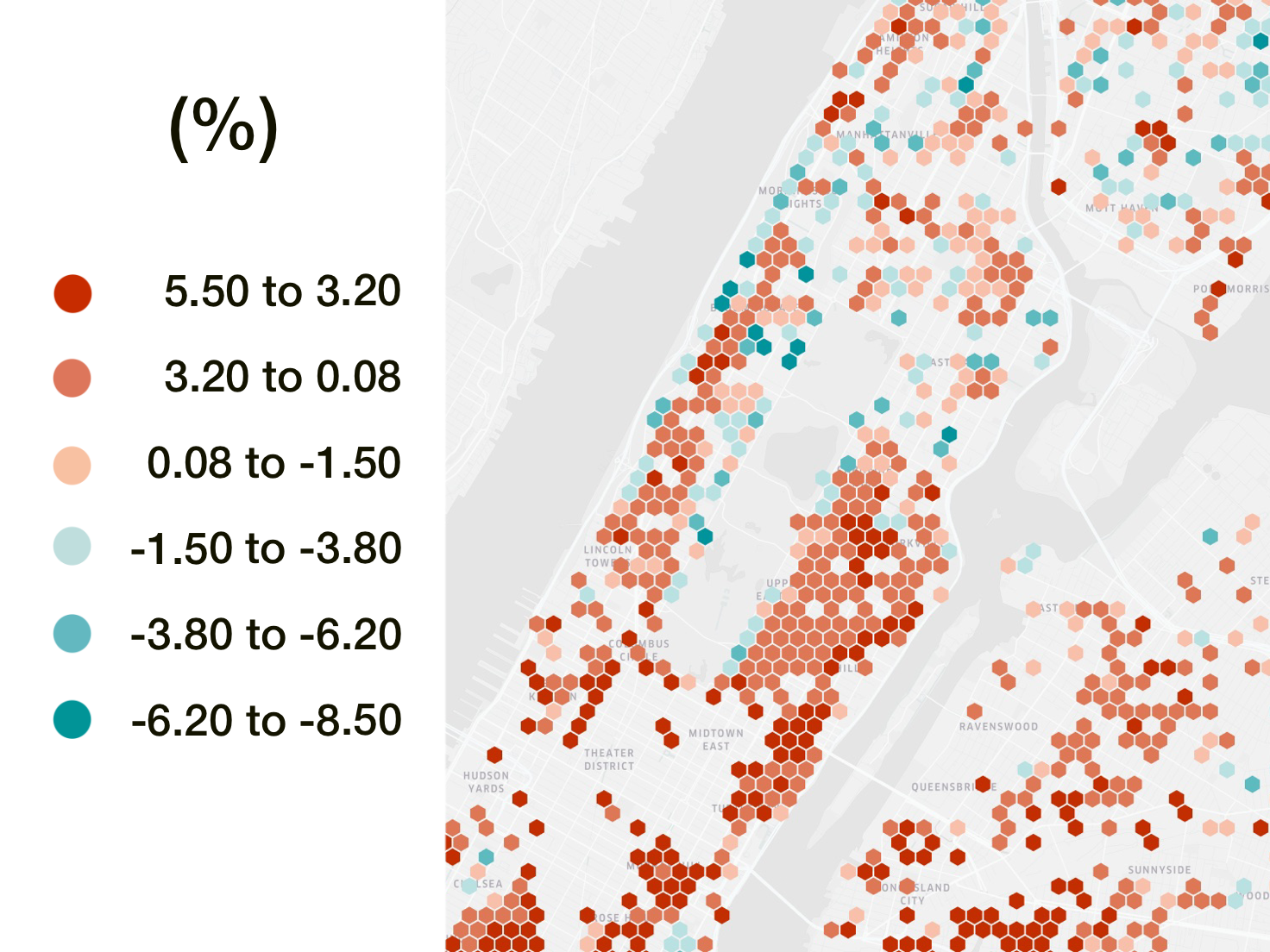}
    \caption{Central Park Effect - NDVI influence as Percent Electricity Change}
    \label{fig:central_park_electric}
\end{figure}

The same maps can be used to explore the regional effects of vegetation on electricity consumption. Within New York City, there are two regional effects which will be highlighted in this analysis to demonstrate the capacity of the system. The first regional effect can be found around central park, as seen in Figure \ref{fig:central_park_electric}. The higher levels of vegetation from the park are estimated to reduce the per unit area electricity consumption of adjacent buildings by 5-7\% compared to their peers in the same neighborhood.

\begin{figure}[h!]
    \centering
    \includegraphics[width=6.5cm]{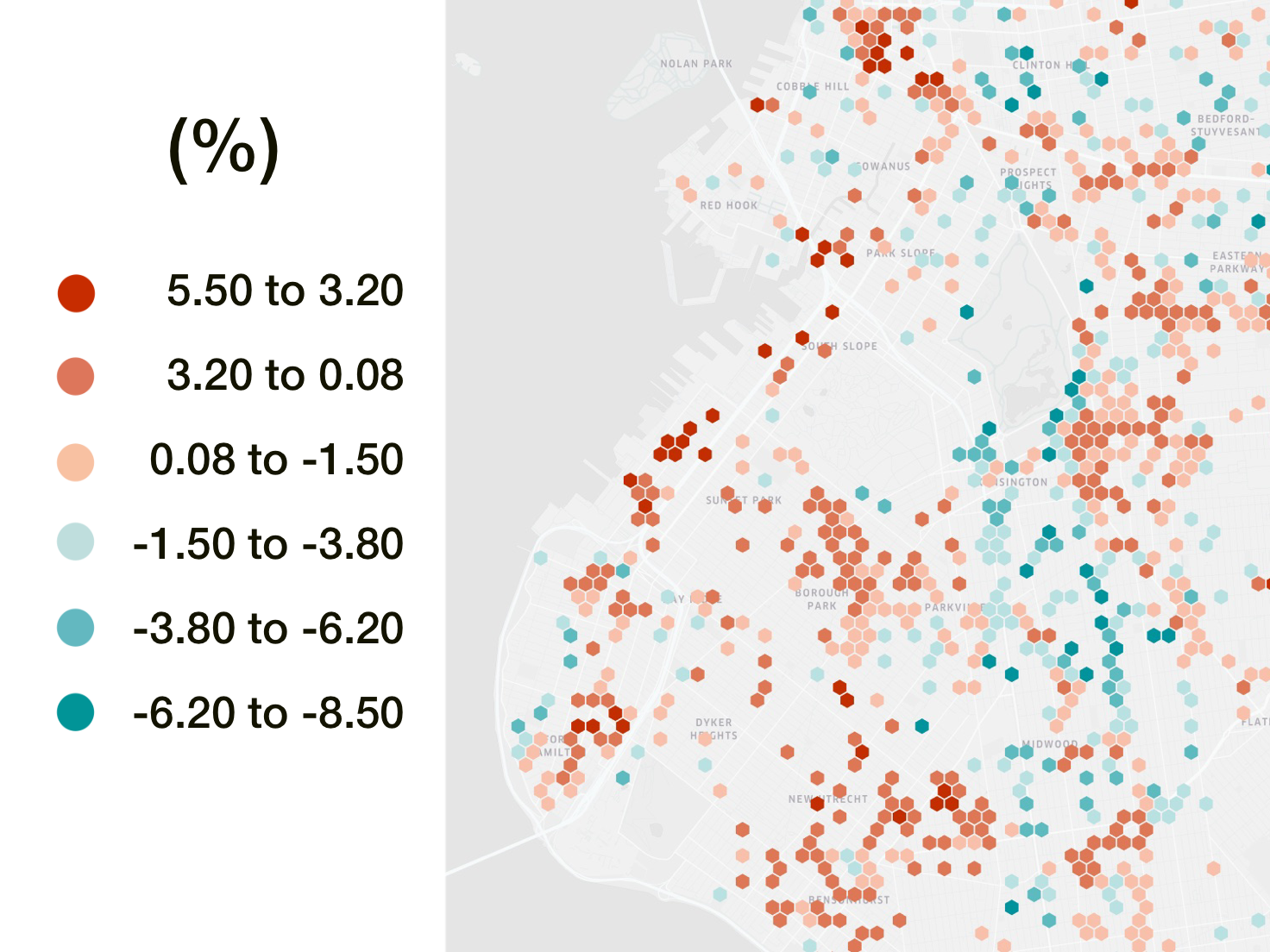}
    \caption{Brooklyn NDVI - Electricity}
    \label{fig:brooklyn_ndvi}
\end{figure}

The second regional effect highlighted in this work can be found in Brooklyn between the streets of Nostrand Ave and Ocean Parkway. With higher relative vegetation, this region is estimated to typically enjoy a 4-6\% reduction electricity per unit area compared to nearby neighborhoods without significant densities of vegetation. As seen in Figure \ref{fig:brooklyn_ndvi}, buildings in the regions directly south of Prospect Park (middle of Brooklyn) are estimated to have a 10\% reduction in electricity consumption per unit area compared to buildings just a few blocks to the southeast.

\subsection{Gas}
The gas data is observed to have much more homogeneous sources of deviation as the regression in Table \ref{tab:regression_mean_centered} is dominated by the  heating degree days, with a coefficient of 0.638. B11 readings from Sentinel-2, capturing shortwave radiation, are again identified as a significant indicator. For the gas regression, we find that higher B11 readings seem to indicate higher gas consumption. As higher B11 readings would indicate that the region is less effective at trapping thermal energy, this matches our intuition.

The summation of the absolute value coefficients for B1, B11, NDVI, WIND, TCDC (cloud coverage), and ACPC01 (precipitation) is 0.49. While heating degree days provide major insights into the behavioral mechanics of gas consumption, nearly half of the picture is missed when environment variables are not included.

In terms of social indicators, we once again see that the night light radiance may provide an important clue for the discrimination of building class. The average radiance coefficient of -0.158 indicates that higher night time lighting may be an indicator for lower gas consumption. It's possible that regions of higher intensity night lighting correspond to commercial zones, which may not require heat throughout the night. Additionally, the total property value of the building seems to serve as a more significant indicator for gas consumption, likely as more expensive buildings may have either transitioned away from gas or again may not have high levels of occupancy at night which require heating.

\subsection{Spatial Analysis - Gas}
\begin{figure}[h!]
    \centering
    \includegraphics[width=6.5cm]{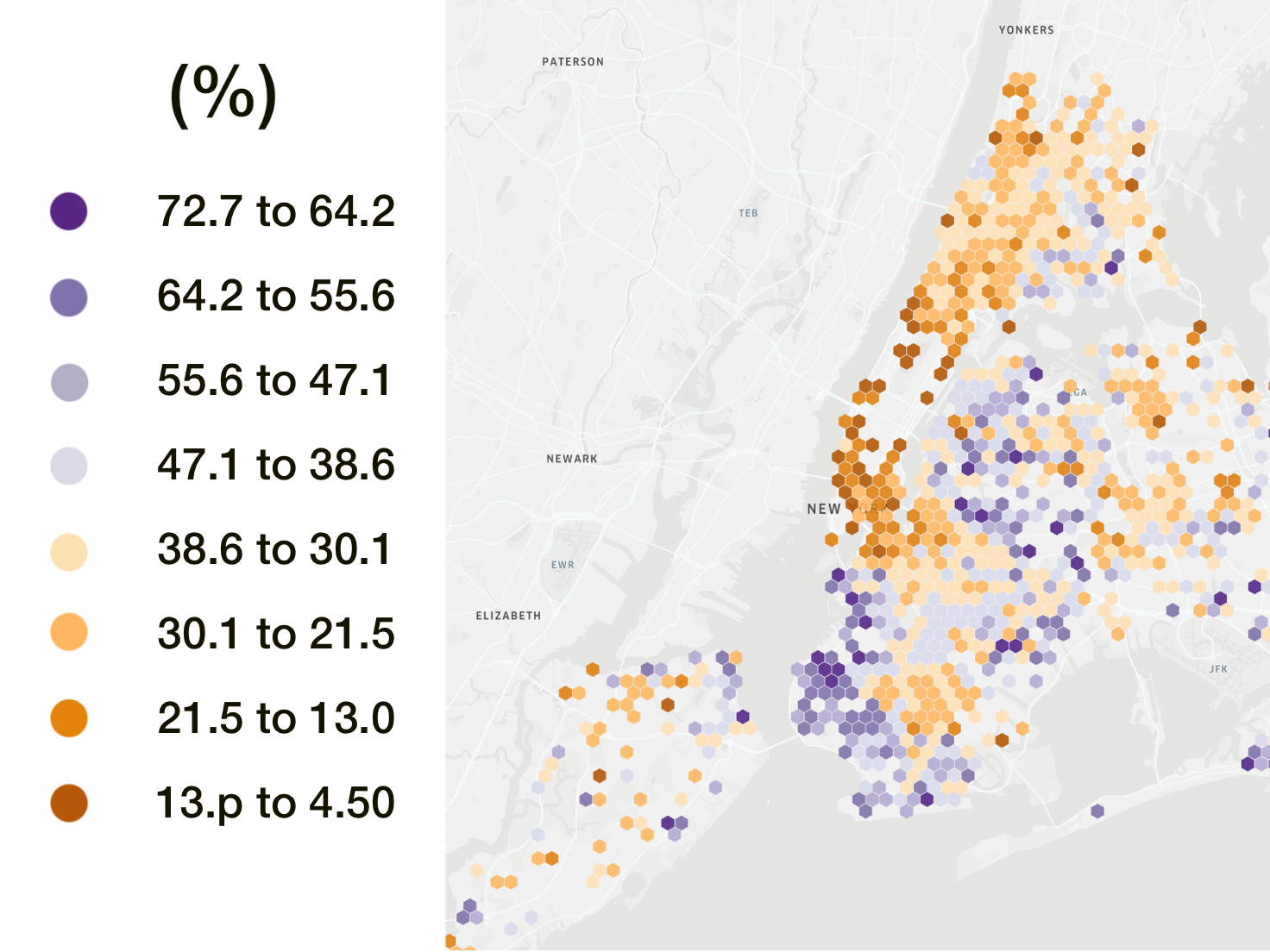}
    \caption{Environmental Percent Difference - Gas}
    \label{fig:environmental_gas}
\end{figure}

Regions in the south of New York City which interface the ocean seemed to have the most significant likely increase in gas consumption, as seen in Figure \ref{fig:environmental_gas}.

The vegetation of the city seems to have almost no appreciable effect on gas consumption. The greenest regions of New York are likely to see a marginal increase of 0.11\% in gas consumption due to vegetation. While trees typically provide a cooling effect in the summer months from evapotranspiration \cite{kleerekoper_how_2012}, they are dormant in the winter months when heating may be required.

\begin{figure}[h!]
    \centering
    \includegraphics[width=6.5cm]{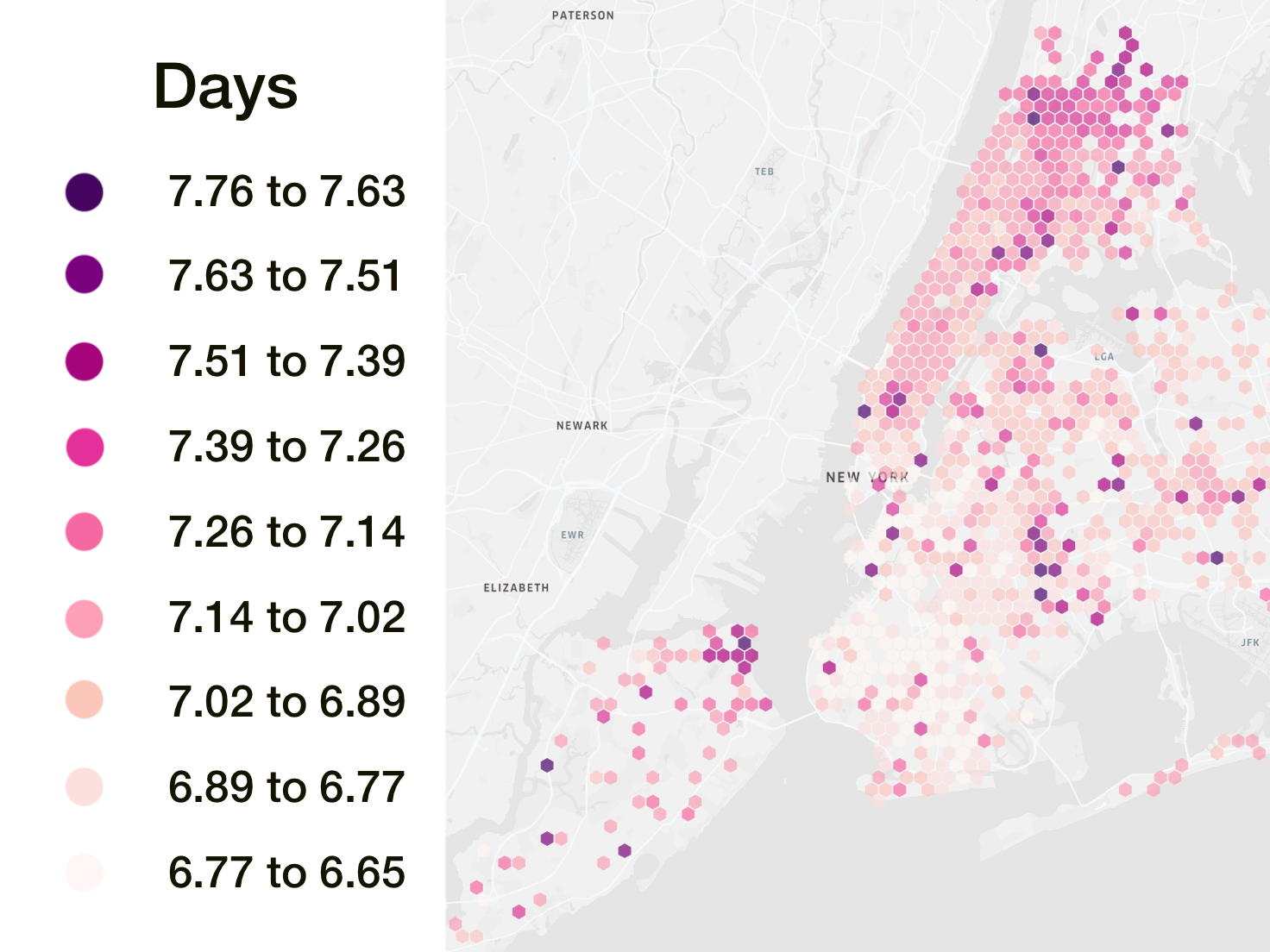}
    \caption{NYC Heating Degree Days}
    \label{fig:heating_degrees}
\end{figure}

The distribution of heating degree days in New York is reasonably flat, with an average monthly spread between 6 and 8 as seen in Figure \ref{fig:heating_degrees}. This indicates that given temperature alone, we might expect the typical building in the Bronx to have a 10\% greater gas consumption than those of Brooklyn. This is not the case however for the full prediction of environmental effects, however, as evidenced by Figure \ref{fig:environmental_gas}.

\begin{figure}[h!]
    \centering
    \includegraphics[width=6.5cm]{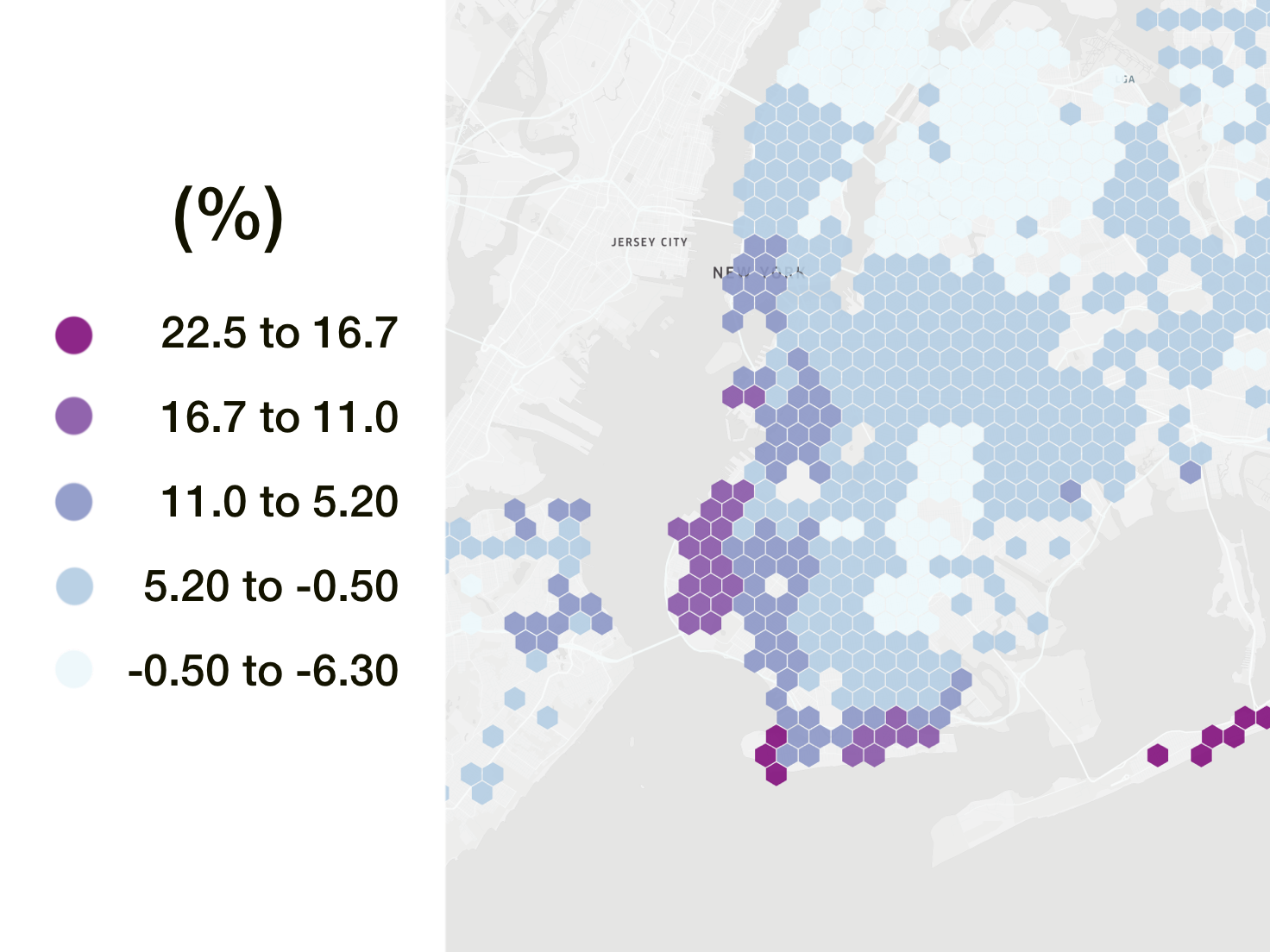}
    \caption{Brooklyn Wind Speeds - Percent Deviation}
    \label{fig:brooklyn_wind}
\end{figure}

Wind has a significant influence on gas consumption in Brooklyn, and management of coastal wind seems to be one of the primary causes of discrepancies with gas consumption. Prior work has demonstrated that the effect of infiltration is more pronounced in older buildings \cite{antretter_effects_2007}. Given that Brooklyn has a high density of older buildings, we might expect wind to have a significant impact on their overall gas consumption. In the coastal regions of southern Brooklyn and Staten Island, wind rivals heating degree days in significance as wind is regularly estimated to increase gas consumption per unit area by 5-15\% as seen in Figure \ref{fig:brooklyn_wind}.

Wind speeds seem to drop off when they interface with larger bodies of vegetation, which likely increase the surface roughness of the urban texture. The reduced wind speeds in Midwood, Flatbush, and Staten Island leads the predictions for environmental effects to hover between a 0 and 15\% increase in gas consumption. This lives in stark contrast to the nearby region of Sunset Park, in which buildings regularly may expect between a 20-30\% increase in gas consumption largely due to increased wind speeds.

Another potentially interesting topic is the role of B11 measurements from Sentinel2. With a positive coefficient, higher readings of B11 indicate the likelihood of a greater quantity of gas consumption per unit area. Higher B11 readings indicate that the region is less effective at naturally trapping thermal energy, which may be detrimental to the building's energy efficiency throughout the cold winter months.

\subsection{Microclimate Transitions}
Microclimate is commonly framed as inducing a consistent environmental impact on a small region of the city. Urban Heat Island, one of the commonly cited forms of urban microclimate, is often portrayed as the consistent addition of heat to a region due to modified heat capacitance, changes in land use, or urban canyon effects \cite{yin_effects_2018}. Often the analysis of microclimate is conducted on the scale of a year, which is likely to still provide valuable information when developing policy for urban decarbonization \cite{reinhart_urban_2016}.

As the data for this analysis is collected with a monthly resolution, we may now explore how a building might transition between environmental microclimates throughout the year. To explore the propensity for microcliamte transitions, the regressed microclimate significance matrix $\textbf{A}$ is clustered into ten unique groups for the purpose of illustration. To maintain consistent terminology, we will use the term "energy microclimate" (EMC) to describe the results of the clustering against $\textbf{A}$. EMCs are perhaps most similar  to Homogeneous Urban Zones (HUZ) which are introduced in the urban land use classification of Madrid \cite{lopez-moreno_identification_2022}. A notable difference between HUZs and EMCs is that EMCs are curated against statistically significant microclimate features conditioned using energy consumption. In practice, EMC groups represent the distinct environmental conditions which have significant influence on the energy consumption of buildings.

Notably, almost every building in New York City will experience transitions through at least eight of the ten potential energy microclimates. This perhaps speaks to the dominance of macroscopic weather patterns, as the energy microclimates are likely to have a significant overlap with seasonal weather patterns in New York City.

\begin{figure}[h!]
  \centering
  \includegraphics[width=6.5cm]{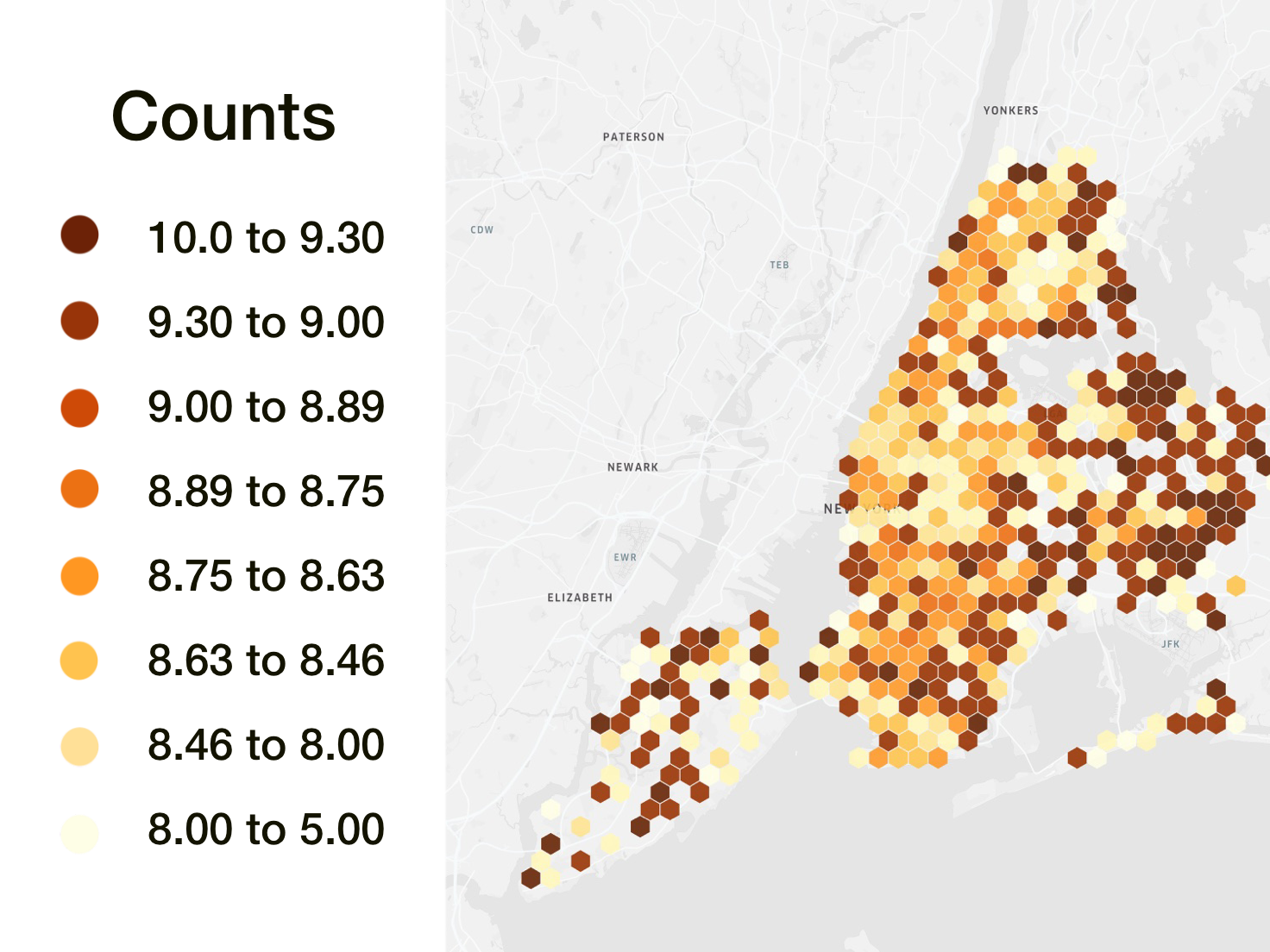}
  \caption{Count of Unique Microclimates}
   \label{fig:microclimate_counts}
\end{figure}

The plot of unique microclimate counts in Figure \ref{fig:microclimate_counts} shows the presence of two potentially significant factors to drive the behavior of New York City's urban climate. The first is the potential presence of an \textit{urban core}, which exists in southern Manhattan and into dips into the western part of Queens. It is likely that the more climatically stable urban core is related to the density of vegetation which has a consistent annual pattern of growth. This theory is validated by the high number of transient regions stretching into Long Island, which is often much greener than the rest of the city.

The second interesting point of note is that of coastal climates which are less likely to participate in auxiliary urban climate dynamics. These regions may be seen on the eastern coast of Staten Island, the southern coast of Brooklyn. While this may give intuition as to the potential transience of urban microclimate in each region of the city, it does not directly indicate greater energy consumption. For example, a building located on the climatically more stable southern cost of Brooklyn may simply experience a more consistent, harsh microclimate. 

\subsection{Electricity Clustering}
For ease of interpretation, the number of cluster zones is reduced to five for all subsequent clustering. In addition, to focus on the effect of micro climate, VIIRS and PLUTO data are removed from the clustering analysis leaving only environmental variables.

\begin{figure}[h!]
  \centering
  \includegraphics[width=7cm]{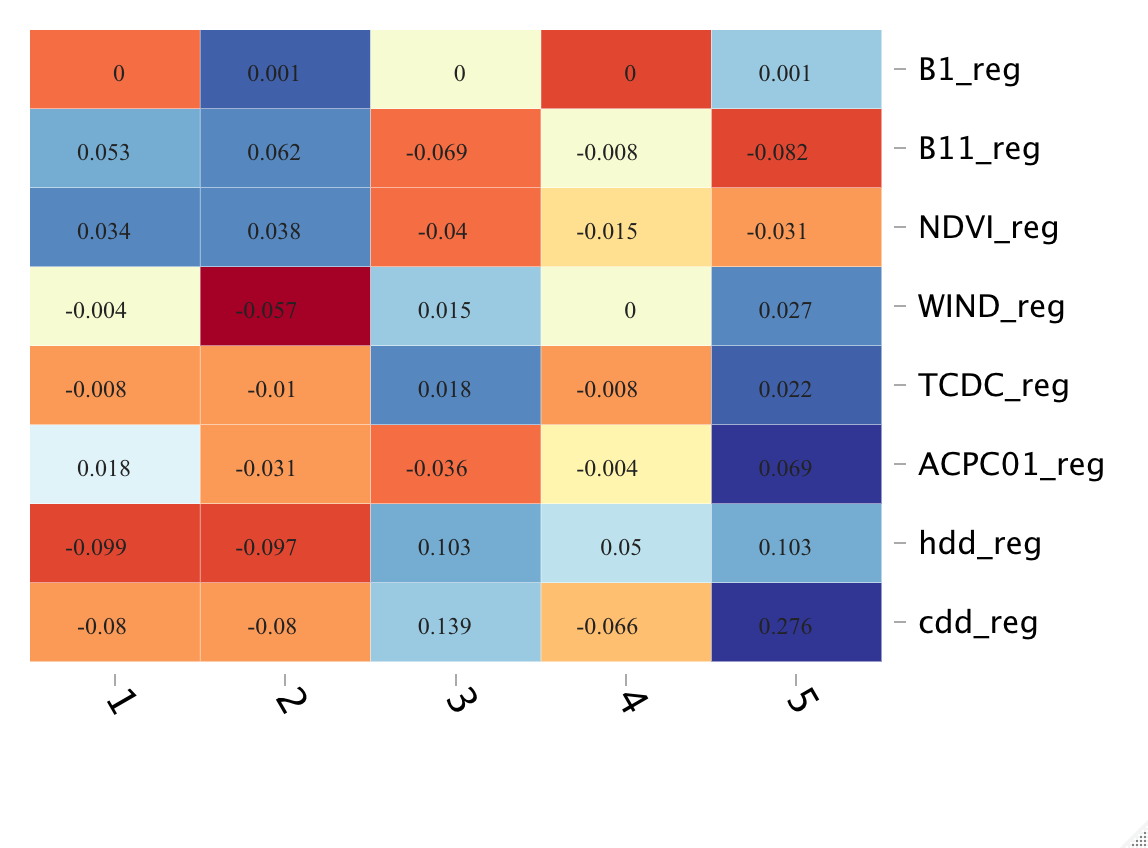}
  \caption{Microclimate Cluster Centroids}
  \label{fig:electric_heatmap}
\end{figure}

The cluster centroids are used to identify the potential behaviors of each cluster, which can be found in the heatmap Figure \ref{fig:electric_heatmap}. The heatmap colors are normalized against the rows, to help identify which terms might be changing between each cluster. The exponential of their summation yields the expected impact of environmental features on electricity consumption, represented as percent deviation.

\begin{figure}[h!]
  \centering
  \includegraphics[width=6.5cm]{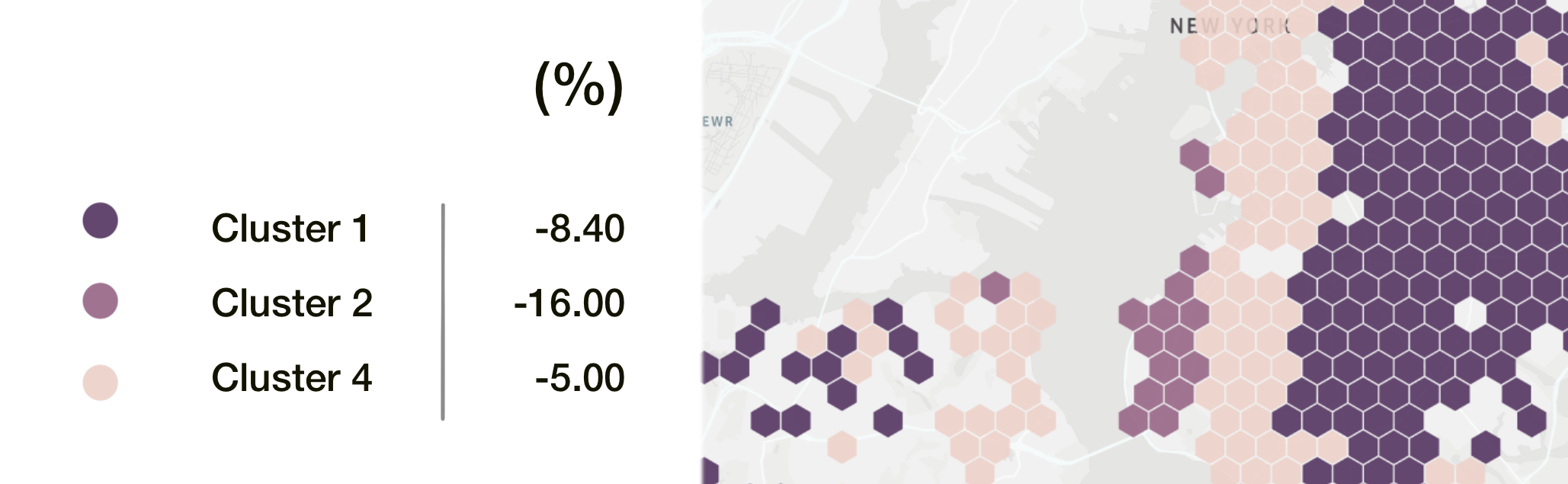}
  \caption{Spring Microclimates - Southern New York City}
  \label{fig:cluster_zones}
\end{figure}

The spring months see a flair of activity for urban microclimates in New York, particularly along the southern coast of the city \ref{fig:cluster_zones}. The coastal climate almost exclusively falls into in cluster two, which is segmented against cluster one almost exclusively for its high wind speeds and cloud coverage.

The spring months have relatively lower requirements for electricity consumption in New York City. As such, the entire city is classified into an EMC with lower than average electricity consumption in the spring.

As electricity is more closely associated with cooling infrastructure in New York City, it matches intuition that electricity demand is likely to be lower in regions with higher wind speeds. Cluster four is likely partitioned due to its higher levels of impervious surfaces, which likely affect the NDVI levels and B11 measurements.


\section{Limitations and Future Work}
While this analysis demonstrates the potential value of using remote sensing systems combined with climate reanalysis, there are some notable limitations regarding the scope of potential applications. As our system was not curated with the primary purpose of predicting building energy consumption, it is likely to miss important nonlinear relationships between variables. Building age and urban night lighting, for example, may host a wealth of information about the construction quality of buildings and their occupancy patterns. However, regions with one fifth the urban nightlighting compared to commercial regions are not necessarily composed of one fifth the commercial regions. Because of this nightlights are likely best expressed through nonlinear relationships with energy consumption.

The relationship between regions in the city is not explicitly incorporated as a mechanism within this model. Our results from the spatial analysis of gas consumption found that high wind speeds were predicted to increase gas consumption on the southern coast of the city. However, as our system cannot predict the change of environmental variables associated with terraforming the region, it is unable to simulate the impact of something like efforts to reduce the wind speed in southern Brooklyn. 

Clustering into energy microclimates is another potential point of interest for the research. While a method was proposed to curate distinct energy microclimates in the city, we recognize that we do not have the capacity to fully simulate the effect of transitioning from one energy microclimate to another due to the coupled nature of environmental parameters. 

Additionally, our system does not claim to provide a mechanism of replacing building energy modeling systems. As peak loads are disregarded in this study, this system has limited potential for grid planning \cite{ang_concept_2020}. Monthly data in New York City was the highest temporal resolution that we could find, and thus remains the limiting temporal factor - we cannot predict with a higher temporal resolution than this.

Bespoke data sources provide a limiting factor to the generalization of our research for new cities. As it stands, we hypothesize that distribution and intensity of urban night lights captured in VIIRS may provide a challenge to generalizing their benefits for large scale urban energy analysis. Additionally building footprints were provided by New York City in a carefully curated data set. Cities without a comprehensive database of building footprints may have limited capacity in generating a comprehensive building energy model of their city. While there seems to be growing potential to automate the task of extracting building footprints \cite{microsoft_usbuildingfootprints_2018}, the current quality of building footprints in dense urban areas is not suitable for analysis.

To our knowledge, energy benchmarking for buildings, a branch of energy related research which seeks to rank buildings against their peers, does not consider the implications of urban microclimate in establishing relational metrics to score buildings. Given that prior work has demonstrated other microclimate phenomena like surface urban heat island (SUHI) may not have equitable impact on all members of society  \cite{hsu_disproportionate_2021}, we believe that the automatic curation of environmental parameters using our system will play a valuable role in promoting more comprehensive metrics for benchmarking research.

Finally, the quantity of building energy consumption data remains a point of concern. Without the capacity to validate the generalization of models, it is a concern that the benefits of microclimate analysis may disproportionately weigh towards those cities which already have a program in place to track and report energy consumption. While remote sensing data makes global microclimate measurements possible, truly bringing the benefits of remote sensing to a global audience will require increased reporting of building energy data worldwide.

\section{Conclusion and Implications}
This study serves as one of the first examples of the direct impact of urban microclimate on building energy consumption based on real-world, historical monthly data over the span of three years. In this study, we demonstrated a method of rapidly extracting hyper-localized environmental features around buildings using a variety of satellite imagery and high resolution reanalysis data. The tools used to capture this data will be made available for all researchers as an open-source tool. This will enable all research related to the built environment to rapidly take a collection of building footprints and capture their historical microclimate data.

This study also demonstrates the potential value of high resolution, historical microclimate data through the use of linear regression. Using the results of the linear regression, we show the significant effects of wind on gas consumption in Brooklyn. We estimate that wind in Brooklyn may increase the gas consumption of buildings in coastal regions by 10\%. We also note the effect of parks on electricity consumption, as buildings directly adjacent to Central Park are predicted to consume 5-7\% less electricity.

Additionally, our study demonstrates the potential value of including urban night lighting as a feature of the regression. Although the VIIRS instrument offers a much lower spatial resolution than that of Sentinel 2 or Landsat8, it provides a valuable source of information as to the use type of buildings in the region of the city. While it's possible that this information is bespoke to New York, the use of VIIRS data for the classification of building type may provide a scalable method of segmenting the behavioral characteristics of buildings by region in cities.

Assuming the removal of outliers by only including prediction results within two standard deviations of the mean, our results indicate that microclimate in urban spaces may play a significant role in driving energy consumption. We show that microclimate may decrease gas consumption by as much as 71\% or increase it by as much as 221\%. Our results indicate that purely environmental effects have the potential to drop the electricity consumption of a typical building in New York City by 24.4\% or increase the electricity consumption by 55.2\%.

In summary, modern research in urban energy consumption has shown urban microclimate to be a significant hurdle for improved accuracy of urban energy modeling. This study demonstrates a mechanism of rapidly collecting high resolution urban microclimate data, utilizing it in a brief case study to explore microclimate effects on energy consumption in New York City. In the process of curating our analysis, we highlight potential engineering challenges associated with cloud coverage, shadows, and seasonal effects. We also explore correlations between environmental microclimate features, using a cleaned version of the data to compose a definition of Energy Microclimates (EMCs). Finally, we show that urban microclimate in New York is expressed through dynamic, competing forces which may have significant localized effects both in time of year and spatial arrangement. The process and data quality laid out in this research may be used to rapidly collect microclimate data for accelerated research research into urban microclimate design strategies. With continued research into urban microclimate and a better understanding of its relationship with energy consumption, microclimate modification may become a more realistic solution to reduce urban scale building energy consumption. In doing so, this research provides a potential pathway to study the invisible walls of microclimate which may help to reduce carbon emissions of our urban spaces.

\section{Acknowledgments}

This work was supported in part by the Precourt Institute for Energy and the U.S. National Science Foundation (NSF) under Grant No. 1941695. Any opinions, findings, and conclusions or recommendations expressed in this material are those of the author(s) and do not necessarily reflect the views of U.S. NSF and/or the Precourt Institute for Energy.

\begin{table*}[!htbp] \centering
  \caption{Linear Model - Mean Centered}
  \label{tab:regression_mean_centered}
\begin{tabular}{@{\extracolsep{5pt}}lcc}
\\[-1.8ex]\hline
\hline \\[-1.8ex]
 & \multicolumn{2}{c}{\textit{Dependent variable:}} \\
\cline{2-3}
\\[-1.8ex] & log( Electric / Area ) & log( Gas / Area ) \\
\\[-1.8ex] & (1) & (2)\\
\hline \\[-1.8ex]
 avg\_rad & 0.002$^{***}$ (0.00004) & $-$0.003$^{***}$ (0.0001) \\
  B1 & 0.060 (0.114) & 3.007$^{***}$ (0.276) \\
  B11 & $-$2.001$^{***}$ (0.061) & 3.971$^{***}$ (0.148) \\
  NDVI & $-$0.754$^{***}$ (0.029) & 0.007 (0.070) \\
  WIND & $-$0.041$^{***}$ (0.003) & 0.096$^{***}$ (0.008) \\
  TCDC & $-$0.002$^{***}$ (0.0003) & 0.006$^{***}$ (0.001) \\
  ACPC01 & 0.783$^{***}$ (0.028) & $-$0.133$^{*}$ (0.069) \\
  hdd & $-$0.015$^{***}$ (0.0004) & 0.097$^{***}$ (0.001) \\
  cdd & 0.047$^{***}$ (0.001) & $-$0.060$^{***}$ (0.003) \\
  elevation & $-$0.003$^{***}$ (0.0001) & $-$0.0001 (0.0003) \\
  assesstot & 0.000$^{***}$ (0.000) & $-$0.000$^{***}$ (0.000) \\
  yearbuilt & 0.001$^{***}$ (0.00003) & 0.0001 (0.0001) \\
  Constant & $-$8.651$^{***}$ (0.002) & $-$8.769$^{***}$ (0.004) \\
 \hline \\[-1.8ex]
Observations & 220,820 & 217,883 \\
R$^{2}$ & 0.116 & 0.171 \\
Adjusted R$^{2}$ & 0.116 & 0.170 \\
Residual Std. Error & 0.748 (df = 220807) & 1.794 (df = 217870) \\
F Statistic & 2,408.626$^{***}$ (df = 12; 220807) & 3,731.927$^{***}$ (df = 12; 217870) \\
\hline
\hline \\[-1.8ex]
\textit{Note:}  & \multicolumn{2}{r}{$^{*}$p$<$0.1; $^{**}$p$<$0.05; $^{***}$p$<$0.01} \\
\end{tabular}
\end{table*}

\begin{table*}[!htbp] 
  \centering
  \caption{Linear Model - Mean Centered \& Normalized}
  \label{tab:regression_normalized}
\begin{tabular}{@{\extracolsep{5pt}}lcc}
\\[-1.8ex]\hline
\hline \\[-1.8ex]
 & \multicolumn{2}{c}{\textit{Dependent variable:}} \\
\cline{2-3}
\\[-1.8ex] & log( Electric / Area ) & log( Gas / Area ) \\
\hline \\[-1.8ex]
 avg\_rad & 0.115$^{***}$ (0.111, 0.119) & $-$0.158$^{***}$ ($-$0.167, $-$0.149) \\
  B1 & 0.001 ($-$0.004, 0.006) & 0.068$^{***}$ (0.055, 0.080) \\
  B11 & $-$0.104$^{***}$ ($-$0.110, $-$0.098) & 0.207$^{***}$ (0.192, 0.222) \\
  NDVI & $-$0.060$^{***}$ ($-$0.064, $-$0.055) & 0.001 ($-$0.010, 0.011) \\
  WIND & $-$0.026$^{***}$ ($-$0.030, $-$0.022) & 0.062$^{***}$ (0.052, 0.071) \\
  TCDC & $-$0.016$^{***}$ ($-$0.021, $-$0.010) & 0.053$^{***}$ (0.040, 0.065) \\
  ACPC01 & 0.055$^{***}$ (0.051, 0.059) & $-$0.009$^{*}$ ($-$0.019, 0.0002) \\
  hdd & $-$0.097$^{***}$ ($-$0.102, $-$0.091) & 0.638$^{***}$ (0.624, 0.651) \\
  cdd & 0.128$^{***}$ (0.122, 0.134) & $-$0.166$^{***}$ ($-$0.180, $-$0.152) \\
  elevation & $-$0.049$^{***}$ ($-$0.053, $-$0.046) & $-$0.002 ($-$0.010, 0.006) \\
  assesstot & 0.073$^{***}$ (0.070, 0.077) & $-$0.256$^{***}$ ($-$0.264, $-$0.248) \\
  yearbuilt & 0.082$^{***}$ (0.079, 0.086) & 0.003 ($-$0.005, 0.012) \\
  Constant & $-$8.651$^{***}$ ($-$8.654, $-$8.648) & $-$8.769$^{***}$ ($-$8.776, $-$8.761) \\
 \hline \\[-1.8ex]
Observations & 220,820 & 217,883 \\
R$^{2}$ & 0.116 & 0.171 \\
Adjusted R$^{2}$ & 0.116 & 0.170 \\
Residual Std. Error & 0.748 (df = 220807) & 1.794 (df = 217870) \\
F Statistic & 2,408.626$^{***}$ (df = 12; 220807) & 3,731.927$^{***}$ (df = 12; 217870) \\
\hline
\hline \\[-1.8ex]
\textit{Note:}  & \multicolumn{2}{r}{$^{*}$p$<$0.1; $^{**}$p$<$0.05; $^{***}$p$<$0.01} \\
\end{tabular}
\end{table*}



\appendix

 \bibliographystyle{acm} 
 \bibliography{main}





\end{document}